# Towards Hydrogen Autarky? Evaluating Import Costs and Domestic Competitiveness in European Energy Strategies

P. Dunkel[1,2,*] T. Klütz[1], J. Linßen[1], D. Stolten[2]

1 Forschungszentrum Jülich GmbH, Institute of Climate and Energy Systems, Jülich Systems Analysis, 52425 Jülich, Germany
2 RWTH Aachen University, Chair for Fuel Cells, Faculty of Mechanical Engineering, 52062 Aachen, German

*corresponding author: p.dunkel@fz-juelich.de

## Abstract

The design of the future European energy system depends heavily on how Europe balances its domestic hydrogen production against its reliance on imports. This study reveals that neither extreme—full self-sufficiency nor complete reliance on imports—is economically optimal through 2050. Using a high-resolution energy system model accounting for interannual weather variability, we find import cost thresholds favoring domestic production decrease from 3.0 €/kg (2030) to 2.5 €/kg (2050). However, the impact of weather is significant and can shift the optimal import share by up to 60 percentage points at constant prices. Strategies with a high import share minimize the need for domestic renewable energy and electrolyzers but require significant investment in long-distance transportation, backup electricity capacity, and storage. Conversely, achieving full self-sufficiency demands massive domestic infrastructure, including up to 1,315 GW of electrolysis by 2050. These findings highlight the critical need for diversified hydrogen strategies that balance cost, resilience, and energy sovereignty. Policy must prioritize flexible infrastructure accommodating both imports and scalable domestic production to navigate evolving market and climatic conditions.

## 1 Introduction

The future hydrogen supply of Europe can follow different strategic pathways. In response to the Russian invasion of Ukraine, the European Union (EU) aims to become more energy independent by 2030, particularly by reducing fossil fuel imports from Russia [1]. As part of this strategy, the European Commission launched the REPowerEU initiative in May 2022, outlining actions for energy savings, diversification of energy sources, and an accelerated expansion of renewable energy. Regarding hydrogen, the EU has set an ambitious target: by 2030, it intends to produce and import 10 million tons of renewable hydrogen each [1].

In the long term, Europe could rely on a mix of imported and domestically produced hydrogen. Imports could arrive via pipelines or ships from outside Europe, while domestic production could be centralized in regions with favorable renewable resources or pursued at the national level to increase autonomy. In an extreme case, Europe could even aim for full hydrogen self-sufficiency to enhance energy security. Each of these strategies would have significant implications for hydrogen infrastructure and system design.

A growing body of literature has evaluated the costs associated with exporting hydrogen and hydrogen-based carriers to Europe, yet results vary widely due to differences in assumptions, system boundaries, and methodological choices. Some studies focus solely on domestic production costs [2,3], while others include transport logistics such as pipelines and shipping to Europe [4–6]. Additionally, approaches differ significantly in how they model renewable energy availability, infrastructure, and the conversion and delivery chain leading to a wide range of cost estimates [7].

Collis and Schomäcker [4] use fixed capacity factors and simple assumptions to estimate hydrogen production and shipping costs globally, identifying Egypt and other regions in North Africa and the Middle East as cost-effective suppliers. Jalbout et al. [5], by contrast, apply hourly renewable energy generation potentials derived from land eligibility assessments to optimize hybrid PV and wind-based systems for export from North Africa via pipelines. Their estimates for hydrogen deliveries to Europe in 2050 range between €1.75 and €2.03/kg. Other studies incorporate more granular energy system modeling. Oyewo et al. [6] use the LUT-ESTM model to simulate hydrogen and derivative exports from Africa, including domestic demand and supply from various renewable sources. They find that pipeline transport from Morocco to Germany in 2050 would cost around €60.2/MWh, compared to €41.4/MWh for local production in Germany. Franzmann et al. [2] derive cost-optimal global hydrogen export cost curves for liquefied hydrogen. Their approach includes a spatially and temporally resolved domestic supply chain—covering onshore wind, open-field PV, electric power transmission, hydrogen pipeline transport, electrolysis, liquefaction, and intermediate storage. The study identifies export costs of below €2.3/kg for an export potential of 79 PWh per year from favorable locations in Africa and the Middle East.

These differences in methodology and assumptions lead to a wide range of cost estimates. For instance, Genge et al. [7] find that hydrogen import costs in 2030 differ by a factor of four across studies, and by a factor of five for 2050. The main sources of this variation are diverging assumptions about capital expenditures and financing conditions. Ship transport cost estimates also vary significantly depending on assumed tanker size, ship costs, and weighted average cost of capital (WACC). In contrast, pipeline import costs from the MENA region show slightly narrower ranges but still vary depending on the use of hourly renewable generation profiles, transport distance assumptions, and infrastructure constraints.

While many studies provide detailed techno-economic assessments of hydrogen export costs, most European energy system models treat hydrogen imports as fixed-cost supply options without representing the underlying infrastructure or cost-driving factors. Imports are often implemented using static cost assumptions and do not account for weather dependent yields, temporal availability, or transport mode-specific characteristics [8–10]. This limits the ability of such models to assess the system-wide implications of different import strategies or cost levels.

Contrary, Lux et al. [11] use the same methodology for import costs as in their European model and consider both pipeline and ship based hydrogen imports. They assume that both pipeline and ship imports are flexibly available at any time. Groß et al. [12] consider pipeline and ship imports. Ship imports are assumed to be constant over the year while for pipeline imports temporally resolved import profiles are used that account for daily and seasonal production variations in the exporting country.

This review highlights the methodological fragmentation in the literature. Assumptions regarding renewable energy potentials, electrolyzer performance, cost of capital, and infrastructure significantly influence the cost outcomes [7,13]. Without harmonized approaches, comparisons across studies remain difficult, and the resulting uncertainty complicates strategic planning for hydrogen imports into Europe. A more consistent and transparent modeling framework is therefore essential to assess the true competitiveness of hydrogen supply chains.

To address this gap and provide policymakers and stakeholders with clearer insights, this study employs a spatially and temporally resolved European energy system model. We systematically analyze the interplay between imported and domestically produced hydrogen across multiple weather years. Specifically, this research seeks to answer the following key questions:

1. What are the cost-competitiveness thresholds for hydrogen imports versus domestic production in Europe, and how do these thresholds evolve towards 2050?

2. How sensitive is the optimal share of hydrogen imports to interannual weather variability within Europe?

3. What are the implications of different hydrogen import shares, ranging from full self-sufficiency to high reliance on imports, for the required scale and type of energy infrastructure within Europe, including renewables, electrolysis, transmission networks, and storage?

4. How do varying shares of hydrogen imports influence overall energy system costs and operational dynamics?

By exploring these questions, this study aims to quantify the trade-offs associated with different hydrogen supply strategies, assessing their impact on infrastructure development, system costs, and resilience under varying market and climatic conditions, thereby contributing to the development of robust and flexible European hydrogen policies.

## 2 Methodology

This section first describes the European energy system model employed in the analysis. Then, the methodology used to derive the hydrogen import cost is outlined. Lastly, the different scenarios are described which are used to address the research questions. Further details regarding the methods can be found in the supplemental information.

### 2.1 Energy System Model

This study investigates the development of hydrogen infrastructure in Europe through 2050 using a spatially and temporally resolved energy system model. A high spatial resolution captures geographic variations in energy production and demand, while a high temporal resolution is essential to reflect fluctuations in renewable generation and demand patterns [14].

The model developed for this study, ETHOS.fineEurope, is based on the open-source Python framework ETHOS.FINE (Framework for Integrated Energy System Analysis) [15]. ETHOS.FINE enables the modeling and optimization of multi-energy carrier systems with flexible spatial and temporal resolutions. The goal of the optimization is to minimize total annual system costs under technical and regulatory constraints. Time series aggregation is applied using the tsam Python package to reduce computational complexity [16].

#### 2.1.1 Model Scope and Assumptions

The model covers the EU-27 countries, as well as the United Kingdom, Norway, and Switzerland. Due to data limitations, other European countries are excluded. The optimization is carried out at ten-year intervals from 2030 to 2050 using a myopic optimization approach.

Spatial resolution follows the NUTS classification [17] (NUTS-0 to NUTS-2), with NUTS-1 as the default to balance resolution and computation time resulting in 101 onshore regions. Additionally, 74 Offshore regions are modeled according to Exclusive Economic Zones (EEZ) to account for offshore wind, gas production and transport.

Each model run covers one year with an hourly resolution. To reduce model complexity, the time series are aggregated to 60 typical days with 12 segments considering a method to be able to capture seasonal storage options [18]. Energy demand from end-use sectors is modeled exogenously with fixed spatial and temporal profiles. The model focuses on networked energy carriers suitable for large-scale transmission—electricity, hydrogen, and natural gas. Ship and truck transport within Europe, including hydrogen derivatives like ammonia, are excluded.

A block diagram of the model structure is shown in Figure 12 in the supplemental information.

#### 2.1.2 Renewable Energy

The model includes onshore and offshore wind power, utility-scale photovoltaic (PV), rooftop PV, and hydropower as renewable energy technologies. Existing capacities of onshore and offshore wind, rooftop and utility-scale PV, and hydropower are incorporated into the model. Hydropower generation profiles and capacities are based on data from 2015 from [19]. Following [20], hydropower expansion is not considered, assuming the technical potential for new large-scale installations in Europe is exhausted. According to IRENA hydropower capacity increased by only 7 GW (5%) in the last 10 years in Europe [21].

Data on existing wind power plants are sourced from The Wind Power database [22] and processed following the methodology of Pena-Sanchez, Dunkel and Winkler et al. [23].

Rooftop and utility-scale PV data are derived from PowerPlantMatching [24], Global Solar Tracker [25], and national sources. Utility-scale PV is assumed to face south, while rooftop PV installations are assumed to be evenly distributed among south, south-east, and south-west orientations. The tilt angle per site is assigned based on optimal values per country using the method from Jacobson and Jadhav [26].

To reflect a recent status of renewable capacity, model-internal capacities are scaled to match IEA national statistics from 2022 [27]. Scaling factors are calculated per country and uniformly applied across regions, assuming that the capacity additions until 2020 followed the distribution of the model-internal capacities. Decommissioning of existing capacity is also included, based on IEA historical capacity trajectories and assumed technical lifetimes. The model includes expansion potentials for onshore wind, offshore wind, rooftop PV, and utility-scale PV. Utility-scale PV capacity potentials are based on [28], offshore wind capacity potentials are taken from [29], while onshore wind potentials are taken from [30]. Rooftop PV potentials rely on Bodis et al. [31], providing a 100x100 m raster for Europe. A surface availability factor of 0.49 and an obstacle reduction factor of 0.6 are applied. Orientation and tilt distributions for rooftop PV potentials are based on Risch et al. [32] using empirical German data. The module LG Electronics LG370Q1C-A5 is used for rooftop PV simulations, and WINAICO WSx-240P6 for utility-scale PV. Investment costs for onshore wind and PV are taken from [33], while costs for offshore wind follow the IEA World Energy Outlook 2023 [34].

Weather-dependent generation time series for PV and Wind are simulated with the tool ETHOS.RESKit using ERA5 reanalysis data following the methodology in Pena-Sanchez, Dunkel and Winkler et al. [23]. Wind turbine power simulation accounts for wake effects and general losses and applies the developed correction of wind capacity factors to the IEA production values. ETHOS.RESKit outputs provide hourly time series per site for onshore and offshore wind as well as rooftop and utility-scale PV for historical weather years. The year 2010, characterized by low wind resource availability, is selected as the default weather year to enable a more robust and resilient system design. [35]. Regional generation profiles and capacities are then aggregated and clustered based on full load hours (FLH). By default, three clusters are used for PV and ten for wind to account for different weather conditions within each region. For existing installations, a single cluster per region and technology is applied.

### 2.1.3 Hydrogen Regasification, Power-to-X and Hydrogen-to-Power

At defined hydrogen import points based on current Liquid Natural Gas (LNG) import terminals [36], liquefied hydrogen ($LH_2$) can be regasified into gaseous hydrogen. These regasification facilities are modeled with unlimited capacity.

For the production of green hydrogen, the model includes Proton Exchange Membrane (PEM) electrolysis as the power-to-hydrogen conversion technology.

Hydrogen-based power generation is represented using open-cycle gas turbines (OCGT), combined-cycle gas turbines (CCGT), and PEM fuel cells. These technologies enable the reconversion of hydrogen into electricity.

### 2.1.4 Conventional Power Plants

All large-scale conventional power plant types available in Europe are represented in the model, including lignite, hard coal, nuclear, gas, and oil-fired power plants. Location and capacity data are sourced from the PowerPlantMatching database [24]. Technology parameters such as efficiency, emissions factors, and typical operational lifetimes are technology-specific but assumed to be location-independent and are taken from [33]. Regional plant capacities are calculated by aggregating all plants within each region.

To reflect climate policy, the model includes annual emission limits per country, in line with the EU's greenhouse gas neutrality target by 2050 and the EU ETS goal of a 62% reduction in emissions by 2030 compared to 2005 levels [37]. Additionally for Germany, the year 2045 is assumed as the national neutrality target year. Emission limits for the years between 2030 and 2050 are linearly interpolated. The operation of fossil-fueled plants is constrained accordingly, plants that emit $CO_2$ may only operate within these national emissions caps.

Due to the long planning and construction times of nuclear plants, new nuclear capacity is assumed to become available only after 2035 and only if already under construction or officially planned, based on Global Energy Monitor data [38]. Given recent developments in the EU, the operational lifetime of existing nuclear plants is extended to 60 years.

To reflect national energy transition plans, phase-out schedules for coal and nuclear power are modeled explicitly. For each country, the officially announced exit years for coal and nuclear power with the status of the year 2022 were researched and incorporated [38–40]. If the model's target year exceeds a country's phase-out year, the corresponding capacities are removed from the system.

### 2.1.5 Energy Storage
**Natural Gas**

Underground storage includes salt caverns and aquifers, with existing capacities sourced from the GIE Storage Database [41] and [42]. Offshore capacities are assigned to the nearest coastal region.

**Hydrogen**

Storage technologies include compressed gas vessels, liquid hydrogen tanks, and geological storage (salt caverns, aquifers). Compressed and liquid hydrogen tanks are unrestricted but only allowed onshore. Geological storage capacity potentials for salt caverns are used from Caglayan et al. [43].

Injection and withdrawal capacities for geological hydrogen storage are constrained by surface-to-underground capacity ratios (1/270 for salt caverns, 1/800 for aquifers). Surface and subsurface capacities are optimized separately. Techno-economic parameters, e.g., cushion gas needs, efficiencies, CAPEX/OPEX, are based on Groß, Dunkel et al. [44]. For the valuation of the cushion gas, a hydrogen price of €2/kg is assumed and taken into account in the investment costs.

**Electricity**

Lithium-ion batteries are available as flexible, unconstrained storage. Pumped hydro and reservoir storage capacities are included based on data from 2015 from [19]. Expansion is not allowed, assuming saturated potential [20].

### 2.1.6 Energy Imports and Production
The model allows unrestricted imports of coal, oil, and uranium at fixed prices, constrained only by emissions limits.

**Natural Gas**

Natural gas can be produced within Europe or imported via LNG terminals or pipelines. Production sites and historical output are derived from [45,46], with pipeline import capacities based on GIE and ENTSOG data [47,48]. LNG import capacities are region-specific based on [36], and all hourly import or production is capped at 1/8760 of the reported annual capacity to prevent unrealistic values.

**Biomass**

Biomass is modeled for both power generation (solid biomass) and biomethane production (liquid biomass). Potentials and costs are derived from the ENSPRESO database (scenario ENS-Med) [49]. Only agricultural residues and forest biomass are included; agricultural crops are excluded due to the potential competition with food production. Municipal waste potential is modeled at the national level and shared across regions of the same country without interregional transport.

### 2.1.7 Transmission Infrastructure

The model includes grid-bound interregional transport of electricity, hydrogen, and natural gas. To keep the computational effort manageable, other modes of transport like trucks and ships are not included. Electricity is transmitted via high-voltage lines and subsea cables, gas and hydrogen via pipelines. Both existing and new infrastructure are modeled.

**Hydrogen and Natural Gas Network**

The model represents the existing European natural gas network using data from the Global Energy Infrastructure dataset [47]. As the model operates with Net Transfer Capacities (NTC), pipeline capacity is derived from pipeline diameters if direct capacity data is not available. This conversion follows the conversion outlined in Gas for Climate [50], which links pipeline diameter to capacity. It is important to note that this method provides a rough estimate, as real pipeline capacity also depends on other factors, such as pipeline length and pressure conditions, which are not captured here.

Additionally, the model allows for the possibility of repurposing existing natural gas pipelines for hydrogen transport. In this case, up to 80% of the gas pipeline's original capacity can be used for hydrogen, based on studies showing that higher hydrogen flow velocities can compensate for its lower energy density compared to natural gas. This conversion follows the methodology from Haeseldonckx et al.[51] and the European Hydrogen Backbone [50].

New pipeline routes, both for natural gas and hydrogen, are considered in the model. It is assumed that pipelines can be built along existing pipeline corridors. Pipeline distances between regions are calculated from the planar Euclidian distance between region centroids. These distances are adjusted with a detour factor of 1.3 to account for realistic routes, based on the work of Welder et al. [52]. In addition to onshore pipelines, offshore hydrogen pipelines can be built along existing natural gas pipeline networks. For these pipelines, the Dijkstra algorithm is used to determine the shortest route and distance between onshore regions.

The techno-economic parameters for both new and repurposed pipelines are based on cost scenarios from European Hydrogen Backbone [53] and Reuss [54]. For hydrogen pipelines, the costs from the "Medium" scenario from the European Hydrogen Backbone are applied, while natural gas pipeline costs are taken from Welder et al. [52]. Additional compressor costs are included to adjust the pressure of hydrogen for pipeline transport, with data sourced from Reuss [54].

**Electricity Grid**

The model includes both the existing high-voltage transmission network and the construction of new power lines. Only NTC are considered as dynamic power flow calculations are not possible due to the model's linear programming formulation.

For determining the transmission capacity and distances between regions, the model utilizes data and algorithms from the PyPSA package [55]. Planned projects from the TYNDP 2022 were added [56]. New power lines can be built between regions that already have an existing electricity grid connection. The distance between regions is calculated by multiplying the distance between region centers by a detour factor of 1.3, based on Welder [52].

Offshore wind farms can be connected to onshore regions via subsea cables. Only regions adjacent to the offshore area and within the same Exclusive Economic Zone (EEZ) are considered for this connection. The distance between onshore and offshore regions is calculated as the shortest straight line from the offshore center to the onshore coastline.

The costs for new power lines are based on real data from the TYNDP 2022 [56], using the median cost of all projects. For onshore lines, the model assumes the construction of overhead lines, as most existing infrastructure is composed of these. Due to a lack of region-specific data, factors like terrain or offshore crossings are not considered.

**Energy Demand**

Energy demand is calculated exogenously using stock models and tools that allow a scenario-based modeling of the development until 2050. No cost optimization was employed.

Due to the limited availability of datasets depicting future energy demand scenarios with high hydrogen shares across all sectors, and the absence of open-source tools for generating such scenarios, this study develops custom model-based demand projections. A detailed description of the methodology can be found in the appendix. Assumed demands for hydrogen, electricity and natural gas can be found in Figure 13 in the appendix.

## 2.2 Import cost modeling

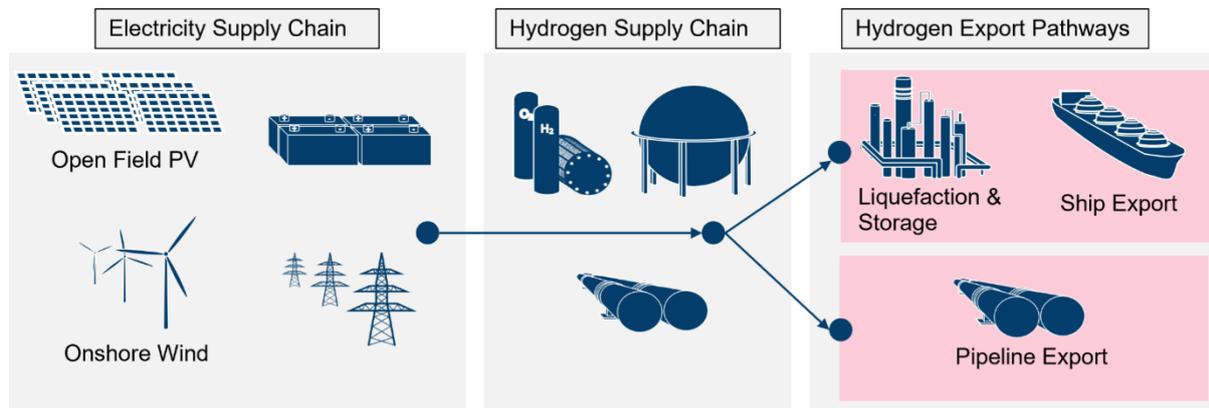

**Figure 1. Modeled supply chain for hydrogen export with ship and pipeline. Adapted from Franzmann et al [2].**

Hydrogen import costs are estimated using a two-stage modeling approach. First, export supply curves are generated by simulating renewable-based hydrogen supply chains in selected exporting countries. Second, transport costs to European import points are added.

**Stage 1: Export Supply Chain Modeling**

The hydrogen supply chains are modeled using a spatially and temporally resolved energy system model based on the ETHOS.FINE framework [15] and implemented with the ModelBuilder Python package [57]. The modeling approach closely follows Franzmann et al. [2], who analyzes hydrogen export supply chains for liquid hydrogen, and Heuser et al. [58], who define global export regions and model infrastructures for liquid hydrogen export. Each supply chain includes onshore wind turbines and open-field photovoltaics (OFPV) as renewable electricity sources. The electricity is used to produce hydrogen via proton exchange membrane (PEM) electrolysis, which is then compressed to 100 bar and transported to export hubs via hydrogen pipelines. Additional infrastructure such as power grids, batteries, hydrogen tanks, and liquefaction units are included where relevant.

Two transport pathways are modeled and depicted in Figure 1: gaseous hydrogen via pipelines and liquid hydrogen via ship transport. While LOHC or ammonia are currently also discussed as potential energy carriers for ship transport, they are not considered in this analysis. Export points are primarily seaports defined by Heuser et al. [58]. For liquid hydrogen, continuous availability at export terminals is ensured through hourly modeling. For gaseous hydrogen, direct transfer to destination via pipeline is assumed, with minimal intermediate storage. Renewable resource potentials are based on Franzmann and Winkler et al. [57]. To account for domestic demand, a maximum of 25% of the available renewable generation potential is assumed for hydrogen exports. Additionally, countries with a hydrogen export potential lower than 2000 TWh/a are discarded for the same reason. Component capacities and locations (e.g., electrolyzers, compressors) are unconstrained, allowing cost-optimal system design. Techno-economic parameters are harmonized with the European model to ensure consistency.

Exporting countries are selected based on preferred regions from Heuser et al. [58], extended to include Turkey. Each country is modeled at the GID-1 administrative level with hourly resolution using a consistent weather year and target year (2030, 2040, or 2050).

Export cost curves are derived for each country up to the 25% export limit, representing the levelized cost of hydrogen (LCOH), calculated as total annualized system costs divided by annual export volume of hydrogen.

**Stage 2: Transport Cost Estimation**

In the second step, transport costs to European import points are added. Shipping costs are based on Johnston et al. [59], with distances calculated using realistic sea routes via the searoute-py Python package [60]. Pipeline transport costs are taken from the Hydrogen Backbone Initiative [53]: 0.08 €/kg/1000 km for repurposed pipelines and 0.16 €/kg/1000 km for new ones. Pipeline routes are estimated based on existing infrastructure and road networks.

European import points include LNG terminals for liquid hydrogen and pipeline entry points in Sicily, Almeria, Gibraltar, and Kiri. While liquid hydrogen can be imported from any modelled export country, pipeline imports are regionally constrained, e.g., Algeria to Sicily and Almeria, Morocco to Gibraltar, Turkey to Kiri.

As cost curves cannot be depicted in the European model for complexity reasons, fixed import costs are assigned per import point. These are based on a simulated import volume of 2000 TWh per year. Export countries unable to meet this volume are excluded to account for their domestic demand. For liquid hydrogen, the average of the three lowest-cost options at the respective import port is used. For pipeline imports, the lowest-cost option per point is selected. Additionally for pipeline imports, hourly export profiles from the source countries are used to define hourly import profiles, capturing seasonal variation in hydrogen availability.

## 2.3 Scenario Description

To analyze the cost competitiveness of domestic hydrogen production within Europe against hydrogen imports and the impact of different hydrogen import shares on the European energy system, two distinct scenarios are created.

For the analysis of the competitiveness of domestic hydrogen production, the model is set up as described in the Section 2.1. No further limitations and constraints are applied. Using the determined hydrogen import costs as described in Section 2.2 as baseline, the model is optimized by varying these import costs between 60% and 130%. For each optimization run, the optimal share of hydrogen import is determined. The runs are conducted for the weather years 1980, 1982, 1983, 1990, 1997, 2002, 2010 and 2017 to capture the effects of different weather conditions within Europe and the countries exporting hydrogen. All runs are conducted by running a myopic optimization capturing the transformation pathway from 2030 to 2050 using 10-year intervals.

In order to compare the impact of different hydrogen import shares on the European energy system, the model is set up with an additional constraint that forces a certain share of hydrogen imports for each model run based on the exogenously defined end-use demand of hydrogen in Europe. Hydrogen demand for reelectrification is part of the optimization results and therefore not included in the constraint. The import share therefore only refers to hydrogen demand in the transport and industry sector. The model is free to utilize the available hydrogen

import technologies, i.e. pipeline and LH2 ship import, to cover the constraint. The import shares are varied between 0 and 100%.. Using the different import share constraints the model is optimized using the weather year 2010. All runs are conducted by running a myopic optimization capturing the transformation pathway from 2030 to 2050 using 10-year intervals.

# 3 Results

In this section, first, the resulting hydrogen import costs for pipeline and liquid hydrogen ship import from different countries to Europe are presented. Afterwards, the competitiveness of European hydrogen production against imports is evaluated by varying import costs. Finally, the impact of different shares of hydrogen imports on the European energy system is assessed.

## 3.1 Hydrogen Import Costs to Europe

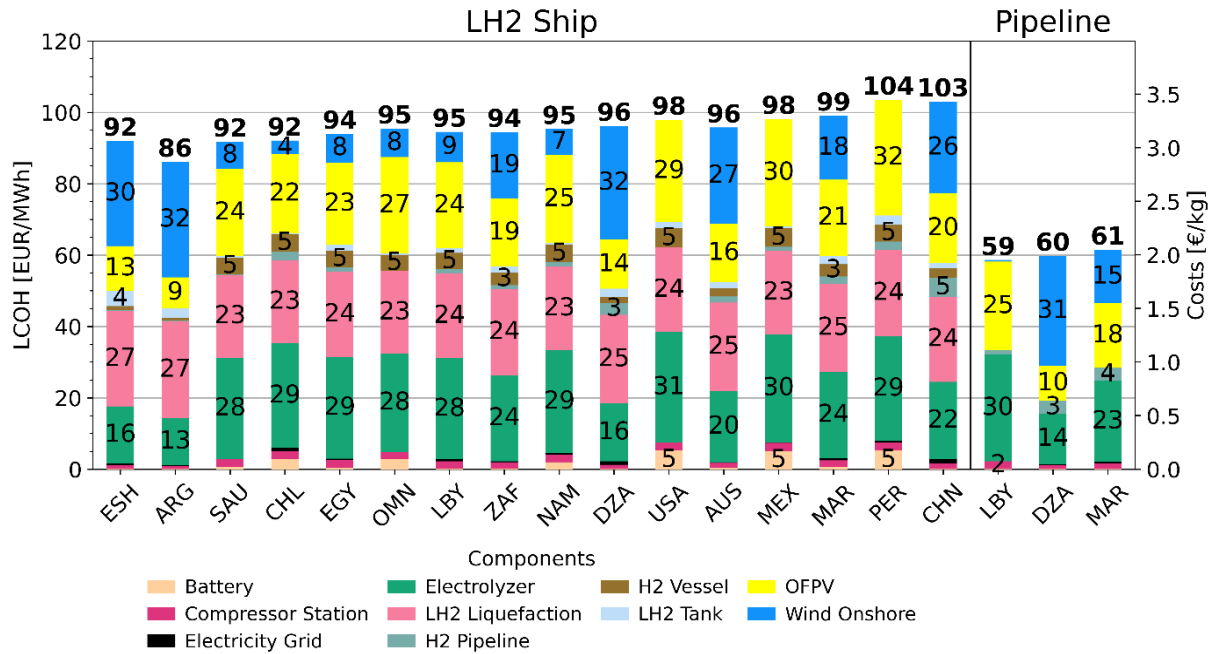

**Figure 2. Composition of hydrogen export costs by export type and country for the weather year 2010 and the target year 2050.**

Figure 2 breaks down hydrogen export costs by technology for the weather year 2010 in the target year 2050 showing cost advantages of pipeline exports over liquid hydrogen exports to Europe. While the lowest liquid hydrogen export cost with 86 €/MWh is observed in Argentina, pipeline export costs are around 59 to 61 €/MWh. The conversion step required for liquid hydrogen, namely liquefaction, results in additional costs of around 24 €/MWh (avg. 25% of the total costs), which do not arise in pipeline exports. For liquid hydrogen exports, the biggest cost drivers besides liquefaction are electrolysis and renewable energy, with average shares of 26 and 38%. Storage and transport costs are in the single-digit percentage range. Furthermore, with a higher share of wind in the system, lower costs for electrolysis can be observed, as is the case in Algeria (DZA) for pipeline exports. This is because, unlike PV, wind is not subject to a day-night cycle, which means that electrolysis can be better utilized. The full-load hours for electrolysis in Algeria are 6,360 hours, while in Libya with high a share of PV they are only 3,020 hours. The fluctuation in export costs for export volumes of 2,000 TWh/a is low, as shown in Figure 11 in the appendix, as the countries considered are far from their potential limits. The calculated export costs are therefore also valid as approximation for other export volumes.

## 3.2 Competitiveness of Domestic Hydrogen Production

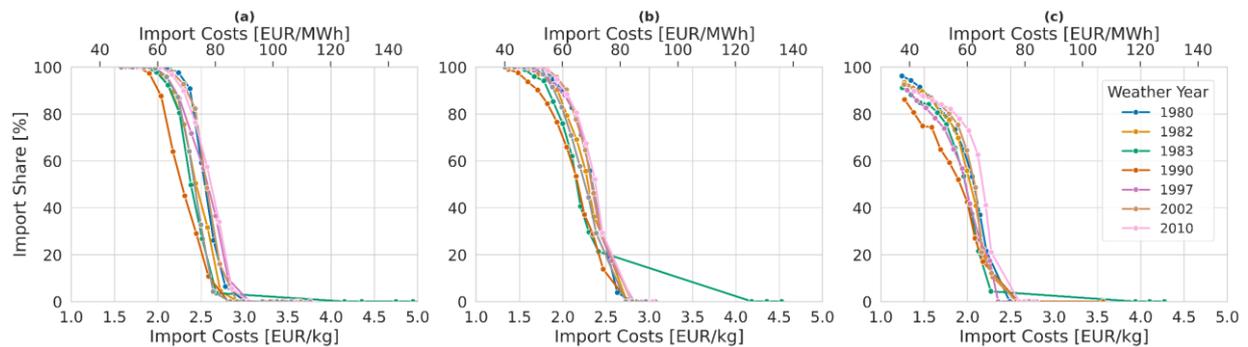

Figure 3. Resulting import shares of hydrogen at different hydrogen import costs for the weather years 1980, 1982, 1983, 1990, 1997, 2002 and 2010 for the target years 2030 (a), 2040 (b) and 2050 (c).

To assess the competitiveness of domestically produced hydrogen, the European energy system is optimized under varying hydrogen import cost scenarios. Figure 3 shows the resulting hydrogen import shares for different import costs in 2030, 2040 and 2050, taking into account eight different weather years. The costs shown in the previous section are used as baseline and varied between 60% and 130%. Within this import cost range, systems with import shares of 0-100% can be found for all weather years. The lowest import quotas are observed for the weather year 1990, and the highest for the weather year 2010 across all target years. This variation directly reflects the impact of weather conditions of renewable energies on domestic production economics: The year 1990 features high onshore wind full-load hours within Europe, significantly lowering the cost of domestic electrolysis and thus boosting the competitiveness of intra-European hydrogen production. Conversely, 2010 represented less favorable meteorological conditions, leading to higher overall energy system costs which increased the cost of domestic renewable electricity; this, in turn, makes hydrogen imports relatively more cost-effective compared to more costly domestic production under those specific conditions. The wide range of cost-optimal import shares in the 2.5-3.0€/kg range for the target year 2030 is also striking. Depending on the weather year, the import share is between 3 and 60% with import costs of around 2.6€/kg. Conversely, with the same import share, cost ranges of up to 0.5€/kg can be observed depending on the weather year. This illustrates the strong influence of weather conditions. Similar observations regarding the range of possible import shares can be made for the following years 2040 and 2050.

With regard to the competitiveness of intra-European hydrogen production, robust import costs of 3.0€/kg can be determined for 2030, above which hydrogen production can be carried out cost-effectively entirely in Europe. In 2040, this figure will be 2.7€/kg and in 2050 2.5 €/kg. A 100% supply of hydrogen to Europe from abroad will be achieved in 2030 for all weather years with import costs below 1.9€/kg, and in 2040 with import costs below 1.5€/kg. This value is not achieved by 2050 within the calculated reference points. It should be noted that liquid hydrogen imports are of no or only minimal relevance in all scenarios, with import volumes of up to a maximum of 17 TWh/a in 2050. Thus, all hydrogen is imported by pipeline from North Africa.

## 3.3 Impacts of Hydrogen Import Shares on Renewable Energy Deployment and Infrastructure in Europe

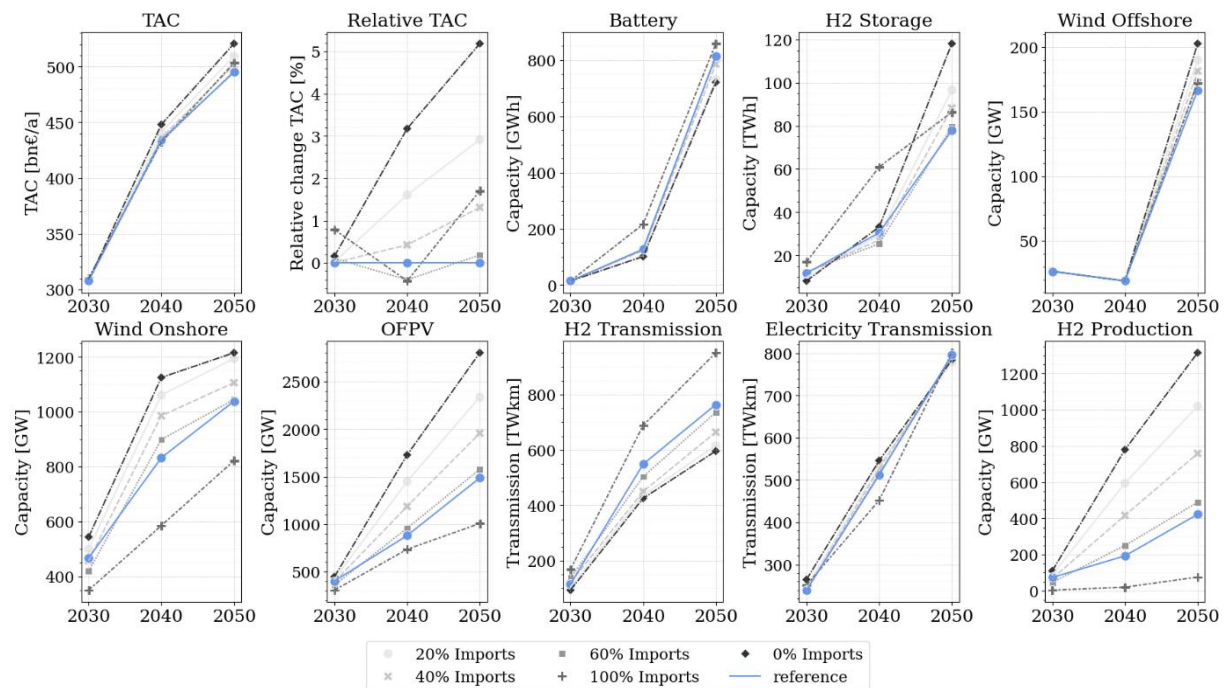

**Figure 4. Development of TAC and capacities of system components from 2030 to 2050 under different import shares.**

The various hydrogen import shares have significant implications for the development of the energy system in Europe. This aspect is evaluated in the following for the weather year 2010. Here the energy system was optimized for different fixed hydrogen import shares.

The required renewable energy (RE) capacity within Europe varies significantly depending on the hydrogen import share. In the reference scenario, which serves as the cost optimal baseline, the cost optimal import share is 34% in 2030, 68% in 2040 and 62% in 2050. Here, 403 GW of PV and 494 GW of Wind are built by 2030. By 2050, RE capacity in Europe increases to 1486 GW for PV and 1205 GW for wind. By enforcing high imports, the demand for renewable energy production is shifting to countries outside Europe.

By 2030, the difference between a 0% and 100% import scenario amounts to 194 GW (+51%) for wind and 141 GW (+26%) for photovoltaics (PV) as shown in Figure 4. The additional PV capacity is primarily installed in Italy (34 GW) and Spain (45.5 GW), with further contributions from France (13 GW), Germany (10.5 GW), the UK (7.4 GW), and Hungary (6.2 GW). All added PV capacity consists of open-field photovoltaic (OFPV) systems, requiring 2820 km² of land assuming 20 km²/GW. Wind capacity expansion is limited to onshore systems; no offshore additions are observed. The main growth occurs in Spain (46 GW), Poland (34 GW), Sweden (28 GW), Norway (24 GW), and France (9 GW), equating to approximately 38,800 new turbines assuming 5 MW per turbine. For comparison, Germany—Europe's leader in onshore wind—had about 29,000 turbines in 2024, with 635 added in 2024.

By 2050, the capacity gap widens to 394 GW for wind onshore (+48%) and 1795 GW for PV (+178%) compared to the full import scenario. Offshore wind grows by 30 GW but remains marginal. Overall, there is no significant spatial shift in RE capacity within Europe, nor major structural changes in the electricity grid when comparing the 0% and 100% import scenario.

### 3.3.1 Impact on Energy System Dynamics

The mix of wind and PV varies with the import strategy. In 2050, wind electricity generation accounts for 49% of RE in the 0% import case and 61% in the 100% import case. This suggests wind is better suited to meet baseload demand, while PV is more closely tied to domestic hydrogen production. When hydrogen is imported, PV capacity declines accordingly, as local electrolysis is no longer required at scale.

This fundamental change in system design also alters how RE is allocated across energy sectors. In the full import case, almost all domestically provided electricity by RE is used to meet electricity demand, while hydrogen end-use demand is entirely satisfied through imports. In contrast, the no-import scenario requires RE to be shared between direct electricity use and hydrogen production, resulting in a much larger overall RE deployment within Europe.

In the 100% import scenario assuming no constraints on grid capacity expansion, about 100 TWh of hydrogen is used for hydrogen-to-power reconversion annually—especially in summer and winter—and primarily in countries such as Spain, the UK, and Italy.. The occurrence of reelectrification under unconstrained conditions signals systemic limitations introduced by reduced domestic RE capacity. The flexibility gap caused by reduced domestic renewable expansion in high-import scenarios is further reflected in increased battery storage requirements. Battery capacity rises from 721 GWh in the 0% import scenario to 859 GWh under 100% import share. The occurrence of reelectrification under unconstrained modeling conditions highlights the potentially valuable role of domestic hydrogen production in reinforcing system flexibility and reliability. This is further reinforced when evaluating the configurations under no additional grid expansion as highlighted in Figure 5. Here, the use of hydrogen reelectrification in Europe increases with increasing shares of hydrogen imports from 111 to 370 TWh (+233%) when no additional grid expansion is permitted. In this case, hydrogen reelectrification occurs mostly in Germany (205 TWh), Belgium (60 TWh) and the UK (30 TWh) during the winter months. This has further impact on the required hydrogen storage capacity which increases by 10% from 204 (0% import share) to 225 TWh (100% import share).

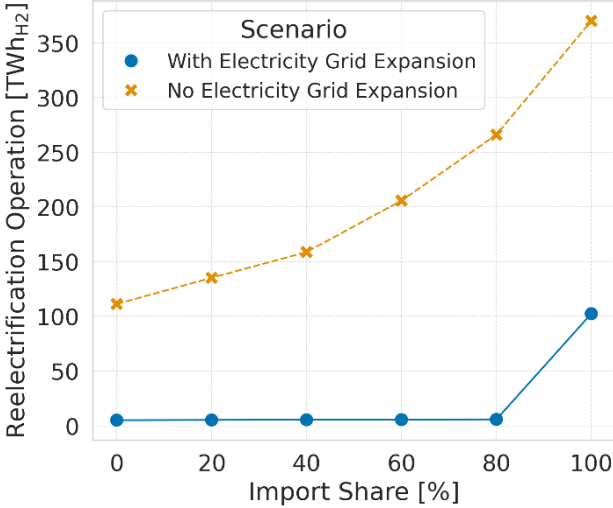

**Figure 5. Hydrogen reelectrification in 2050 under different import shares with and without additional electricity grid expansion.**

By enabling the expansion of renewable capacity beyond the immediate needs of the power sector, domestic hydrogen production creates a dual benefit: excess renewable electricity capacity can support electricity supply during periods of high demand or low generation. During these periods, hydrogen production can shut down and supply can be covered by storage. This is illustrated in Figure 6 that compares the electricity balance for Western Poland for a 0% and 100% import scenario without grid expansion. In the 0% import scenario, electrolysis plants are shut down during periods of low electricity generation from wind such that electricity demand can be covered, while in the 100% import scenario this is not feasible due to the significantly reduced renewable capacity. This interlinkage strengthens resilience against volatility in electricity generation and demand. This underscores the extent to which domestic hydrogen production can serve as a flexibility anchor, reducing the system's reliance on costly or limited short-term balancing resources.

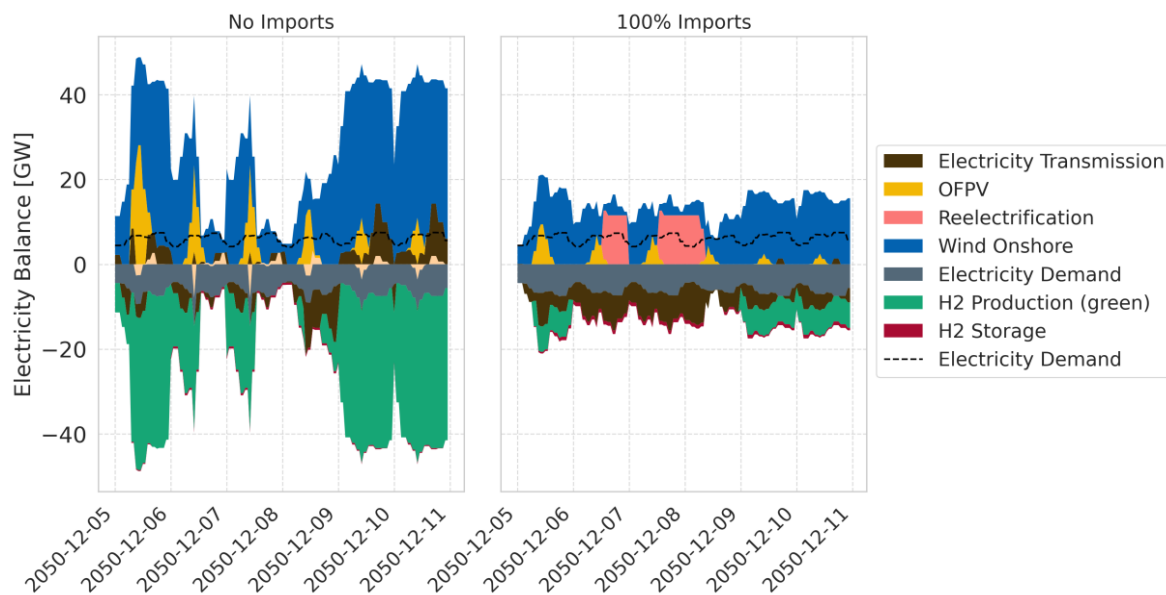

**Figure 6. Electricity balance for western Poland in winter for the 0% and 100% import scenario in 2050 without no additional electricity grid expansion.**

In summary, while hydrogen imports reduce the need for domestic RE and electrolysis capacity , they also limit the ability of the energy system to respond to fluctuations in electricity demand. In contrast, domestic hydrogen production, although more resource-intensive, provides critical redundancy and flexibility by additional renewable electricity capacity that can support electricity supply during periods of high demand.

### 3.3.2   Infrastructure Implications for Hydrogen Supply

Electric grid capacity remains relatively stable across scenarios, ranging from 784 to 800 TWkm by 2050. In contrast, hydrogen infrastructure shows strong sensitivity to import shares. Hydrogen storage needs range from 78 to 118 TWh and decline with higher imports. Pipeline capacity increases from 94–169 TWkm in 2030 to 596–951 TWkm by 2050, peaking under full import scenarios due to long transport distances.

Electrolyzer capacity varies most significantly: from 76 GW (100% import) to 1315 GW (0% import) in 2050. Import capacity at Europe's external borders ranges up to 637 GW. When replacing imports with domestic electrolyzers, Spain, Norway, and Poland are initially prioritized. However, only Spain and Italy show substantial capacity increases at higher self-sufficiency levels, where hydrogen is produced from PV-based electricity. Two main reasons explain this trend: First, by 2050, PV sees a stronger cost decline than wind, making PV-based hydrogen production more attractive, as already observed in the previous section. Second, the cost-effective wind energy potential in Norway and Poland becomes largely exhausted, limiting their capacity expansion.

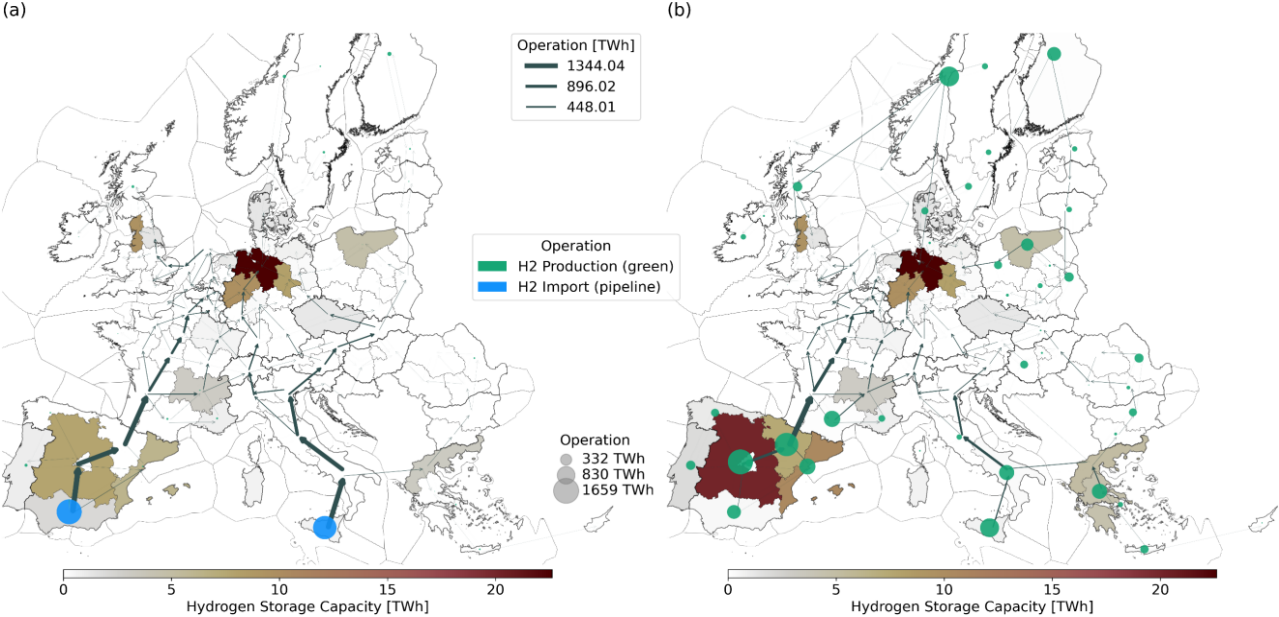

**Figure 7. Hydrogen storage, transport and production at 100% hydrogen import share (a) and 0% import share (b) in the year 2050.**

### 3.3.3 Spatial and Temporal Effects of Hydrogen Storage

Figure 7 illustrate how hydrogen infrastructure scales spatially. Major import corridors from Spain and Italy to Central Europe are significantly larger in the 100% import scenario. Conversely, a 0% import scenario requires less transport capacity, as hydrogen is produced closer to demand locations, reducing transit distances.

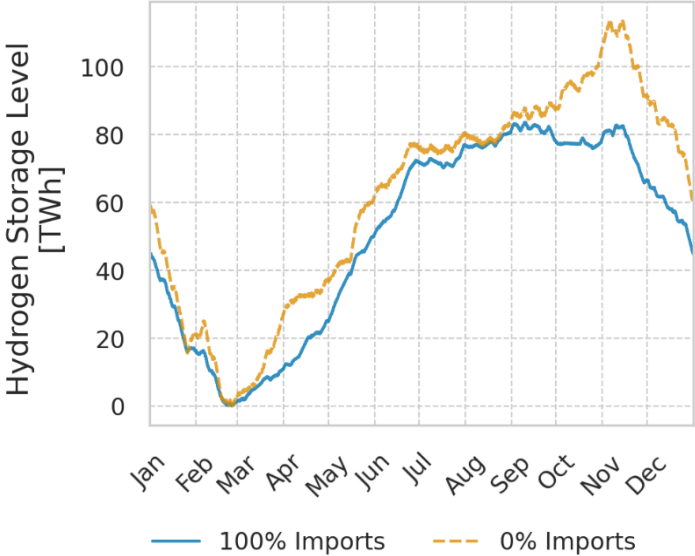

**Figure 8. Storage levels for the 0% and 100% import scenario in the target year 2050.**

Hydrogen storage distribution shows minor spatial variation except for Spain, but differences in seasonal fill levels are evident as shown in Figure 8. In 2050, Spain exhibits lower storage levels under the 100% import scenario and minimal change from July to November. In contrast, the 0% import scenario shows storage buildup during this period and a marked depletion from November to March—indicating a higher need for seasonal buffering. This is due to the stronger seasonal variation in Spanish PV output compared to more stable North African PV production.

Germany shows similar trends. Although total cavern capacity remains unchanged, storage dynamics differ: the 0% import scenario features plateau phases in summer and fall, followed by a sharp rise in November. These patterns align with seasonal wind variations—low in summer, increasing in autumn.

### 3.3.4 System Cost Implications and Strategic Considerations

Neither complete self-sufficiency nor full hydrogen imports are cost optimal. A fully domestic hydrogen supply increases system costs by at least 5% (based on weather year 2010). Since constraints were not applied, this value likely represents a lower bound. Compared to the 100% import scenario, achieving 0% imports would require at least 424 GW additional wind and 1780 GW PV by 2050, plus over 1300 GW of electrolyzer capacity—demanding substantial acceleration of current expansion rates.

On the other hand, full reliance on imports reduces domestic RE and electrolyzer capacity, limiting synergies between electricity and hydrogen production and increasing dependency on alternative flexibility solutions such as batteries, grid expansion, and reelectrification.

Hydrogen production in North Africa offers more predictable output due to lower seasonal variability, reducing the need for large-scale storage and enhancing system planning reliability. A robust European hydrogen strategy must balance cost, security, and flexibility trade-offs between domestic production and imports.

# 4 Conclusion and Discussion

This study explores the competitiveness of domestic versus imported hydrogen production in Europe, considering weather variability, system costs, and infrastructure implications through 2050 by leveraging a high-resolution energy system model.

First and foremost, the findings clearly demonstrate that a hybrid hydrogen strategy is cost-optimal for Europe through 2050. Neither complete hydrogen self-sufficiency nor full reliance on imports represents the most economical pathway.

The competitiveness of intra-European hydrogen production is strongly weather-dependent. For instance, import cost thresholds for favoring domestic production decline over time—from approximately 3.0 €/kg in 2030 to 2.5 €/kg in 2050—indicating improving cost parity. However, year-to-year weather variations can shift the cost-optimal import share by over 50 percentage points at the same price level, highlighting the sensitivity of hydrogen system design to renewable variability.

The strategic choice regarding the import share also fundamentally shapes Europe's energy infrastructure needs, creating distinct investment profiles. High levels of hydrogen imports substantially reduce the need for domestic renewable generation by up to 424 GW for wind and 1800 GW of PV capacity and electrolyzers but increase demand for long-distance hydrogen transport, hydrogen reelectrification capacity, and energy storage. Conversely, domestic production strengthens synergies between electricity and hydrogen sectors but requires significant infrastructure scaling—especially for electrolysis and renewables. For example, a 0% import scenario in 2050 demands over 1300 GW of electrolyzers, a figure well beyond current deployment trajectories.

System flexibility emerges as a key differentiator between these strategies. Relying heavily on imports diminishes the intrinsic flexibility offered by a large, integrated domestic renewable and electrolysis system. The considerable hydrogen reelectrification capacity found to be optimal in the 100% import scenario highlights that imports primarily fulfill bulk energy requirements but may not efficiently address temporal and spatial power balancing needs without dedicated flexibility assets. Domestic hydrogen production, despite its resource intensity, inherently fosters a larger renewable energy fleet capable of contributing more directly to electricity grid stability and resilience.

These findings underscore the importance of system flexibility and diversified supply strategies. Hydrogen imports, especially from North Africa, offer more consistent output and reduce seasonal storage needs. Yet they also pose geopolitical and infrastructure dependencies. In contrast, domestic production reinforces energy sovereignty but is capital-intensive and highly sensitive to interannual weather patterns.

Compared to previous studies, this work provides a more granular spatial and temporal assessment, incorporating hourly data across multiple weather years. Several publications find a self-sufficient hydrogen supply for Europe to be feasible [20,61,62]. Göke et al. [63] find optimal import shares of 0.2%, Host and Klann [64] find that 2-5% of the hydrogen demand is met by imports from North Africa. Furthermore, [65] find that an intra-European hydrogen production is cost competitive and imports are only needed in case of restricted renewable energy expansion. Similar observations are made by [10] that find cost optimal import shares of 1%. They also find that hydrogen reelectrification gained in importance with the increasing share of imports,. Conversely, Seck et al. [66], Frischmuth et al. [67] and Fleiter et al. [8] find much higher hydrogen import shares for 2050 of up to 15%, 37% and 10%. Similarly to this study, Kountouris et al. [68] find that a self-sufficient hydrogen supply results in 2.77% higher system costs, whereas this study finds a value of 5%. No study highlights the strong impact of weather conditions on optimal hydrogen import shares. The diverging results highlight the relevance of consistent assumptions and the choice of methodology when determining hydrogen import costs and resulting optimal hydrogen import shares for Europe.

From a policy perspective, the results highlight the urgency of ramping up both domestic production capacity and strategic import infrastructure. Relying solely on domestic supply would require unprecedented acceleration in wind, solar, and electrolyzer deployment. Meanwhile, over-reliance on imports could expose the system to external shocks and weaken the coupling between renewable and hydrogen sectors. EU hydrogen policy must therefore balance cost optimization with resilience, flexibility, and strategic autonomy. In the first place, Europe needs a clear hydrogen strategy and build the required infrastructure accordingly.

This analysis does not account for certain real-world constraints such as geopolitical risk, public acceptance of infrastructure, or regulatory bottlenecks. Future research should integrate stochastic modeling of geopolitical disruptions, region specific discount rates that capture various risks, deeper techno-economic assessments of alternative hydrogen carriers as well as the option of blue hydrogen, and models that capture dynamic interactions between policy and system development—where policy decisions adapt in response to system outcomes, and in turn influence future deployment pathways.

In conclusion, Europe's hydrogen future is unlikely to be binary. Optimal solutions will combine domestic and imported hydrogen in regionally tailored mixes, with infrastructure built to enable flexibility rather than lock-in. Policymakers and planners should aim for a modular, resilient hydrogen network that can evolve with market conditions, technology costs, and geopolitical developments.

## 4.1 Acknowledgments

Part of this work has been carried out within the framework of the HyUSPRe project which has received funding from the Fuel Cells and Hydrogen 2 Joint Undertaking (now Clean Hydrogen Partnership) under grant agreement No 101006632. This Joint Undertaking receives support from the European Union's Horizon 2020 research and innovation programme, Hydrogen Europe and Hydrogen Europe Research.

This work was supported by the Helmholtz Association under the program "Energy System Design".

## 4.2 Authors contributions

**Philipp Dunkel**: Writing – original draft, Visualization, Validation, Methodology, Investigation, Formal analysis, Conceptualization. **Theresa Klütz**: Writing – review & editing, Supervision, Funding acquisition, Conceptualization. **Jochen Linßen**: Writing – review & editing, Supervision. **Detlef Stolten**: Supervision, Funding acquisition.

All authors have read and agreed to the published version of the manuscript.

### 4.3 Declaration of generative AI and AI-assisted technologies in the manuscript preparation process

During the preparation of this work the authors used ChatGPT (GPT-4) and DeepL Write in order to improve language, clarity and structure. After using these tools, the authors reviewed and edited the content as needed and take full responsibility for the content of the published article.

# 5 References

[1] COMMUNICATION FROM THE COMMISSION TO THE EUROPEAN PARLIAMENT, THE EUROPEAN COUNCIL, THE COUNCIL, THE EUROPEAN ECONOMIC AND SOCIAL COMMITTEE AND THE COMMITTEE OF THE REGIONS REPowerEU Plan. 2022.

[2] Franzmann D, Heinrichs H, Lippkau F, Addanki T, Winkler C, Buchenberg P, et al. Green hydrogen cost-potentials for global trade. International Journal of Hydrogen Energy 2023;48:33062–76. https://doi.org/10.1016/j.ijhydene.2023.05.012.

[3] Fraunhofer ISE. POWER-TO-X COUNTRY ANALYSES - Site-specific, comparative analysis for suitable Power-to-X pathways and products in developing and emerging countries. Fraunhofer Institute for Solar Energy Systems ISE; 2023.

[4] Collis J, Schomäcker R. Determining the Production and Transport Cost for H2 on a Global Scale. Front Energy Res 2022;10. https://doi.org/10.3389/fenrg.2022.909298.

[5] Jalbout E, Genge L, Muesgens F. H2Europe: an analysis of long-term hydrogen import-potentials from the MENA region. 2022 18th International Conference on the European Energy Market (EEM), Ljubljana, Slovenia: IEEE; 2022, p. 1–7. https://doi.org/10.1109/EEM54602.2022.9921055.

[6] Oyewo AS, Lopez G, ElSayed M, Galimova T, Breyer C. Power-to-X Economy: Green e-hydrogen, e-fuels, e-chemicals, and e-materials opportunities in Africa. Energy Reports 2024;12:2026–48. https://doi.org/10.1016/j.egyr.2024.08.011.

[7] Genge L, Scheller F, Müsgens F. Supply costs of green chemical energy carriers at the European border: A meta-analysis. International Journal of Hydrogen Energy 2023. https://doi.org/10.1016/j.ijhydene.2023.06.180.

[8] Fleiter T, Fragoso J, Lux B, Alibaş Ş, Al-Dabbas K, Manz P, et al. Hydrogen Infrastructure in the Future $CO_2$-Neutral European Energy System—How Does the Demand for Hydrogen Affect the Need for Infrastructure? Energy Tech 2024;13:2300981. https://doi.org/10.1002/ente.202300981.

[9] Göke L, Kendziorski M, Kemfert C, Hirschhausen CV. Accounting for spatiality of renewables and storage in transmission planning. Energy Economics 2022;113:106190. https://doi.org/10.1016/j.eneco.2022.106190.

[10] Lux B, Gegenheimer J, Franke K, Sensfuß F, Pfluger B. Supply curves of electricity-based gaseous fuels in the MENA region. Computers & Industrial Engineering 2021;162:107647. https://doi.org/10.1016/j.cie.2021.107647.

[11] Lux B, Frömel M, Resch G, Hasengst F, Sensfuß F. Effects of different renewable electricity diffusion paths and restricted european cooperation on Europe's hydrogen supply. Energy Strategy Reviews 2024;56:101589. https://doi.org/10.1016/j.esr.2024.101589.

[12] Groß T, Dunkel P, Linßen J, Stolten D. EU-scale hydrogen system scenarios. H2020 HyUSPRe project report; 2024.

[13] Schuler J, Ardone A, Fichtner W. A review of shipping cost projections for hydrogen-based energy carriers. International Journal of Hydrogen Energy 2024;49:1497–508. https://doi.org/10.1016/j.ijhydene.2023.10.004.

[14] Kigle S, Mohr S, Kneiske T, Clees T, Ebner M, Harper R, et al. TransHyDE-Sys: An Integrated Systemic Approach for Analyzing and Supporting the Transformation of Energy Systems and Hydrogen Infrastructure Development. Energy Technology 2024;n/a:2300828. https://doi.org/10.1002/ente.202300828.

[15] Klütz T, Knosala K, Behrens J, Maier R, Hoffmann M, Pflugradt N, et al. ETHOS.FINE: A Framework for Integrated Energy System Assessment. Journal of Open Source Software 2025;10:6274. https://doi.org/10.21105/joss.06274.

[16] Hoffmann M, Kotzur L, Stolten D. The Pareto-optimal temporal aggregation of energy system models. Applied Energy 2022;315:119029. https://doi.org/10.1016/j.apenergy.2022.119029.

[17] Eurostat. NUTS - Nomenclature of territorial units for statistics n.d.

[18] Kotzur L, Markewitz P, Robinius M, Stolten D. Time series aggregation for energy system design: Modeling seasonal storage. Applied Energy 2018;213:123–35. https://doi.org/10.1016/j.apenergy.2018.01.023.
[19] Syranidou C. Integration of Renewable Energy Sources into the Future European Power System Using a Verified Dispatch Model with HighSpatiotemporal Resolution. Dr. Forschungszentrum Jülich GmbH Zentralbibliothek, Verlag, 2020.
[20] Pickering B, Lombardi F, Pfenninger S. Diversity of options to eliminate fossil fuels and reach carbon neutrality across the entire European energy system. Joule 2022;6:1253–76. https://doi.org/10.1016/j.joule.2022.05.009.
[21] IRENA. Renewable capacity statistics 2024. Abu Dhabi: International Renewable Energy Agency; 2024.
[22] The wind power. World wind farms database 2023.
[23] Peña-Sánchez EU, Dunkel P, Winkler C, Heinrichs H, Prinz F, Weinand J, et al. Towards high resolution, validated and open global wind power assessments 2025. https://doi.org/10.48550/arXiv.2501.07937.
[24] Gotzens F, Heinrichs H, Hörsch J, Hofmann F. Performing energy modelling exercises in a transparent way - The issue of data quality in power plant databases. Energy Strategy Reviews 2019;23:1–12. https://doi.org/10.1016/j.esr.2018.11.004.
[25] Global Energy Monitor. Global Solar Power Tracker 2022.
[26] Jacobson MZ, Jadhav V. World estimates of PV optimal tilt angles and ratios of sunlight incident upon tilted and tracked PV panels relative to horizontal panels. Solar Energy 2018;169:55–66. https://doi.org/10.1016/j.solener.2018.04.030.
[27] Renewable Energy Progress Tracker – Data Tools. IEA 2024. https://www.iea.org/data-and-statistics/data-tools/renewables-data-explorer (accessed March 25, 2024).
[28] Ryberg DS. Generation lulls from the future potential of wind and solar energy in Europe n.d.:433.
[29] Caglayan DG, Ryberg DS, Heinrichs H, Linßen J, Stolten D, Robinius M. The techno-economic potential of offshore wind energy with optimized future turbine designs in Europe. Applied Energy 2019;255:113794. https://doi.org/10.1016/j.apenergy.2019.113794.
[30] IEA. Global Hydrogen Review 2024 2024.
[31] Bódis K, Kougias I, Jäger-Waldau A, Taylor N, Szabó S. A high-resolution geospatial assessment of the rooftop solar photovoltaic potential in the European Union. Renewable and Sustainable Energy Reviews 2019;114:109309. https://doi.org/10.1016/j.rser.2019.109309.
[32] Risch S, Maier R, Du J, Pflugradt N, Stenzel P, Kotzur L, et al. Potentials of Renewable Energy Sources in Germany and the Influence of Land Use Datasets. Energies 2022;15:5536. https://doi.org/10.3390/en15155536.
[33] Stolten D, Kullmann F, Schöb T, Maier R, Freitag P, Schulze K, et al. Energieperspektiven 2030 - Welche Maßnahmen zur Transformation des deutschen Energiesystems müssen bis 2030 umgesetzt werden, was muss bis 2030 vorbereitet sein? 2023.
[34] IEA. World Energy Outlook 2023. Paris: IEA; 2023.
[35] Grochowicz A, van Greevenbroek K, Benth FE, Zeyringer M. Intersecting near-optimal spaces: European power systems with more resilience to weather variability. Energy Economics 2023;118:106496. https://doi.org/10.1016/j.eneco.2022.106496.
[36] Global Energy Monitor. Global Gas Infrastructure Tracker 2022.
[37] UNFCCC. GHG data from UNFCCC 2019.
[38] Global Energy Monitor. Global Nuclear Power Tracker 2022.
[39] Coal Exit Tracker. Europe Beyond Coal n.d. https://beyond-coal.eu/coal-exit-tracker/ (accessed November 2, 2021).
[40] Global Energy Monitor. Global Coal Plant Tracker 2022.
[41] GIE. GIE Storage Database 2021.
[42] Cavanagh, AJ, Yousefi, SH, Wilkinson, M, Groenenberg, RM. Hydrogen storage potential of existing European gas storage sites in depleted gas fields and aquifers. H2020 HyUSPRe project report; 2022.

[43] Caglayan DG, Weber N, Heinrichs HU, Linßen J, Robinius M, Kukla PA, et al. Technical potential of salt caverns for hydrogen storage in Europe. International Journal of Hydrogen Energy 2020;45:6793–805. https://doi.org/10.1016/j.ijhydene.2019.12.161.
[44] de Maigret J, Viesi D. Report on equipment requirements and capital as well as operating costs for the hydrogen scenarios n.d.
[45] Global Energy Monitor. Global Oil and Gas Extraction Tracker 2022.
[46] Eurostat. Supply, transformation and consumption of gas 2022. https://doi.org/10.2908/NRG_CB_GAS.
[47] Global Energy Infrastructure. Global Gas Pipelines 2022.
[48] European Gas Flow dashboard by ENTSOG n.d. https://gasdashboard.entsog.eu/ (accessed August 17, 2022).
[49] Ruiz P, Nijs W, Tarvydas D, Sgobbi A, Zucker A, Pilli R, et al. ENSPRESO - an open, EU-28 wide, transparent and coherent database of wind, solar and biomass energy potentials. Energy Strategy Reviews 2019;26:100379. https://doi.org/10.1016/j.esr.2019.100379.
[50] Gas for Climate. European Hydrogen Backbone - how a dedicated hydrogen infrastructure can be created. 2020.
[51] Haeseldonckx D, D'haeseleer W. The use of the natural-gas pipeline infrastructure for hydrogen transport in a changing market structure. International Journal of Hydrogen Energy 2007;32:1381–6. https://doi.org/10.1016/j.ijhydene.2006.10.018.
[52] Welder L. Optimizing Cross-linked Infrastructure for Future Energy Systems. Dissertation. Forschungszentrum Jülich GmbH Zentralbibliothek Verlag, 2022.
[53] Gas for Climate. Extending the European Hydrogen Backbone - A European Hydrogen Infrastructure Vision Covering 21 Countries. 2021.
[54] Reuss M. Techno-ökonomische Analyse alternativer Wasserstoffinfrastruktur. Dissertation. Forschungszentrum Jülich GmbH, 2019.
[55] Brown T, Victoria M, Zeyen E, Hofmann F, Neumann F, Frysztacki M, et al. PyPSA-Eur: An open sector-coupled optimisation model of the European energy system 2025.
[56] ENTSO-E, ENTSOG. TYNDP 2022 - Scenario Report. 2022.
[57] Franzmann D, Winkler C, Dunkel P, Stargardt M, Burdack A, Ishmam S, et al. Impact of Spatial Aggregation on Global Renewable Energy System with ETHOS.modelBuilder 2025. https://doi.org/10.2139/ssrn.5252316.
[58] Heuser P-M, Grube T, Heinrichs H, Robinius M, Stolten D. Worldwide Hydrogen Provision Scheme Based on Renewable Energy 2020:27.
[59] Johnston C, Ali Khan MH, Amal R, Daiyan R, MacGill I. Shipping the sunshine: An open-source model for costing renewable hydrogen transport from Australia. International Journal of Hydrogen Energy 2022;47:20362–77. https://doi.org/10.1016/j.ijhydene.2022.04.156.
[60] Gent. genthalili/searoute-py 2024.
[61] Neumann F, Zeyen E, Victoria M, Brown T. The potential role of a hydrogen network in Europe. Joule 2023. https://doi.org/10.1016/j.joule.2023.06.016.
[62] Wetzel M, Gils HC, Bertsch V. Green energy carriers and energy sovereignty in a climate neutral European energy system. Renewable Energy 2023;210:591–603. https://doi.org/10.1016/j.renene.2023.04.015.
[63] Göke L, Weibezahn J, Kendziorski M. How flexible electrification can integrate fluctuating renewables. Energy 2023;278:127832. https://doi.org/10.1016/j.energy.2023.127832.
[64] Horst J, Klann U. MENA-Fuels Analyse eines globalen Marktes für Wasserstoff und synthetische Energieträger hinsichtlich künftiger Handelsbeziehungen 2022.
[65] European Commission. Directorate General for Energy., Fraunhofer Institute for Systems and Innovation Research. METIS 3, study S5: the impact of industry transition on a CO2 neutral European energy system. LU: Publications Office; 2023.
[66] Seck GS, Hache E, Sabathier J, Guedes F, Reigstad GA, Straus J, et al. Hydrogen and the decarbonization of the energy system in europe in 2050: A detailed model-based analysis. Renewable and Sustainable Energy Reviews 2022;167:112779. https://doi.org/10.1016/j.rser.2022.112779.

[67] Frischmuth F, Berghoff M, Braun M, Härtel P. Quantifying seasonal hydrogen storage demands under cost and market uptake uncertainties in energy system transformation pathways. Applied Energy 2024;375:123991. https://doi.org/10.1016/j.apenergy.2024.123991.
[68] Kountouris I, Bramstoft R, Madsen T, Gea-Bermúdez J, Münster M, Keles D. A unified European hydrogen infrastructure planning to support the rapid scale-up of hydrogen production. Nat Commun 2024;15:5517. https://doi.org/10.1038/s41467-024-49867-w.
[69] Tabula Project Team. TABULA_FinalReport. Darmstadt: IWU Institut Wohnen und Umwelt; 2012.
[70] European Commission. Joint Research Centre. JRC-IDEES: Integrated Database of the European Energy Sector : methodological note. LU: Publications Office; 2017.
[71] Simon Pezzutto, Stefano Zambotti, Silvia Croce, Pietro Zambelli. Hotmaps Project, D2.3 WP2 Report – Open Data Set for the EU28 2019.
[72] Sandberg NH, Sartori I, Heidrich O, Dawson R, Dascalaki E, Dimitriou S, et al. Dynamic building stock modelling: Application to 11 European countries to support the energy efficiency and retrofit ambitions of the EU. Energy and Buildings 2016;132:26–38. https://doi.org/10.1016/j.enbuild.2016.05.100.
[73] WorldPop :: Population Counts 2020. https://hub.worldpop.org/project/categories?id=3 (accessed April 15, 2025).
[74] Eurostat. Employment by sex, age, full-time/part-time, professional status and NUTS 2 region 2025. https://doi.org/10.2908/LFST_R_LFE2EFTPT.
[75] Regionalisation of the Final Energy Consumption of the Sector Services in Europe – opendata.ffe.de 2020. https://opendata.ffe.de/regionalisation-of-the-final-energy-consumption-of-services-in-europe/ (accessed April 15, 2025).
[76] Ruhnau O, Hirth L, Praktiknjo A. Time series of heat demand and heat pump efficiency for energy system modeling. Sci Data 2019;6:189. https://doi.org/10.1038/s41597-019-0199-y.
[77] Marlon Schlemminger. ML_household_end-use_load-profiles 2021. https://doi.org/10.25835/0043305.
[78] oemof/demandlib 2025.
[79] Groß T, Dunkel P, Franzmann D, Heinrichs H, Linßen J, Stolten D. Report on H2 supply from Renewable Energy Sources, H2 demand centers and H2 transport infrastructure. H2020 HyUSPRe Project Report 2022:67 pp incl. appendix.
[80] Eurostat. Stock of vehicles by category and NUTS 2 region 2025. https://doi.org/10.2908/TRAN_R_VEHST.
[81] Eurostat. Road freight transport by region of loading (t, tkm, journeys) - annual data 2022. https://doi.org/10.2908/ROAD_GO_TA_RL.
[82] Eurostat. Passengers embarked and disembarked in all ports by direction - annual data 2022. https://doi.org/10.2908/MAR_PA_AA.
[83] Eurostat. Freight and mail air transport routes between partner airports and main airports in Germany 2022. https://doi.org/10.2908/AVIA_GOR_DE.
[84] Eurostat. Gross weight of goods handled in all ports by direction - annual data 2022. https://doi.org/10.2908/MAR_GO_AA.
[85] Ortiz-Imedio R, Caglayan DG, Ortiz A, Heinrichs H, Robinius M, Stolten D, et al. Power-to-Ships: Future electricity and hydrogen demands for shipping on the Atlantic coast of Europe in 2050. Energy 2021;228:120660. https://doi.org/10.1016/j.energy.2021.120660.
[86] Khalili S, Rantanen E, Bogdanov D, Breyer C. Global Transportation Demand Development with Impacts on the Energy Demand and Greenhouse Gas Emissions in a Climate-Constrained World. Energies 2019;12:3870. https://doi.org/10.3390/en12203870.
[87] Eurostat. Railway passengers transported 2022. https://doi.org/10.2908/RAIL_PA_TOTAL.
[88] Eurostat. Railway goods transported 2023. https://doi.org/10.2908/RAIL_GO_TOTAL.
[89] EuroGeographics. Euro Global Map 2022.

[90] European Commission. Directorate General for Energy., European Commission. Directorate General for Climate Action., European Commission. Directorate General for Mobility and Transport. EU reference scenario 2020: energy, transport and GHG emissions : trends to 2050. LU: Publications Office; 2021.
[91] Gas For Climate, Wang A, Jens J, Mavins D, Moultak M, Schimmel M, et al. EUROPEAN HYDROGEN BACKBONE - Analysing future demand, supply, and transport of hydrogen. Guidehouse; 2021.
[92] DNV. Maritime Forecast to 2050. 2021.
[93] Schemme S, Breuer JL, Köller M, Meschede S, Walman F, Samsun RC, et al. H2-based synthetic fuels: A techno-economic comparison of alcohol, ether and hydrocarbon production. International Journal of Hydrogen Energy 2020;45:5395–414. https://doi.org/10.1016/j.ijhydene.2019.05.028.
[94] Bazzanella A, Ausfelder F. Low carbon energy and feedstock for the European chemical industry: Technology Study. DECHEMA, Gesellschaft für Chemische Technik und Biotechnologie eV; 2017.
[95] Directive 2003/87/EC of the European Parliament and of the Council of 13 October 2003 establishing a system for greenhouse gas emission allowance trading within the Union and amending Council Directive 96/61/EC (Text with EEA relevance). 2024.
[96] European Commission. Joint Research Centre. Energy efficiency and GHG emissions: prospective scenarios for the chemical and petrochemical industry. LU: Publications Office; 2017.
[97] Eurostat. PRODCOM. statistics by products 2020.
[98] Rehfeldt M, Fleiter T, Toro F. A bottom-up estimation of the heating and cooling demand in European industry. Energy Efficiency 2018;11:1057–82. https://doi.org/10.1007/s12053-017-9571-y.
[99] Guminski A. CO2 Abatement in the European Industry Sector - Evaluation of Scenario-Based Transformation Pathways and Technical Abatement Measures. Technische Universität München, 2021.
[100] European Commission. Joint Research Centre. The POTEnCIA central scenario: an EU energy outlook to 2050. LU: Publications Office; 2019.
[101] Abrell J. Database for the european union transaction log. Database 2021.
[102] Bundesamt für Umwelt. Schadstoffregister SwissPRTR 2024.
[103] Global Energy Monitor. Global Steel Plant Tracker 2022.
[104] Eurofer. Where is steel made in Europe? 2021. https://www.eurofer.eu/about-steel/learn-about-steel/where-is-steel-made-in-europe/ (accessed October 5, 2021).

# 6 Appendix

## 6.1 Hydrogen Imports

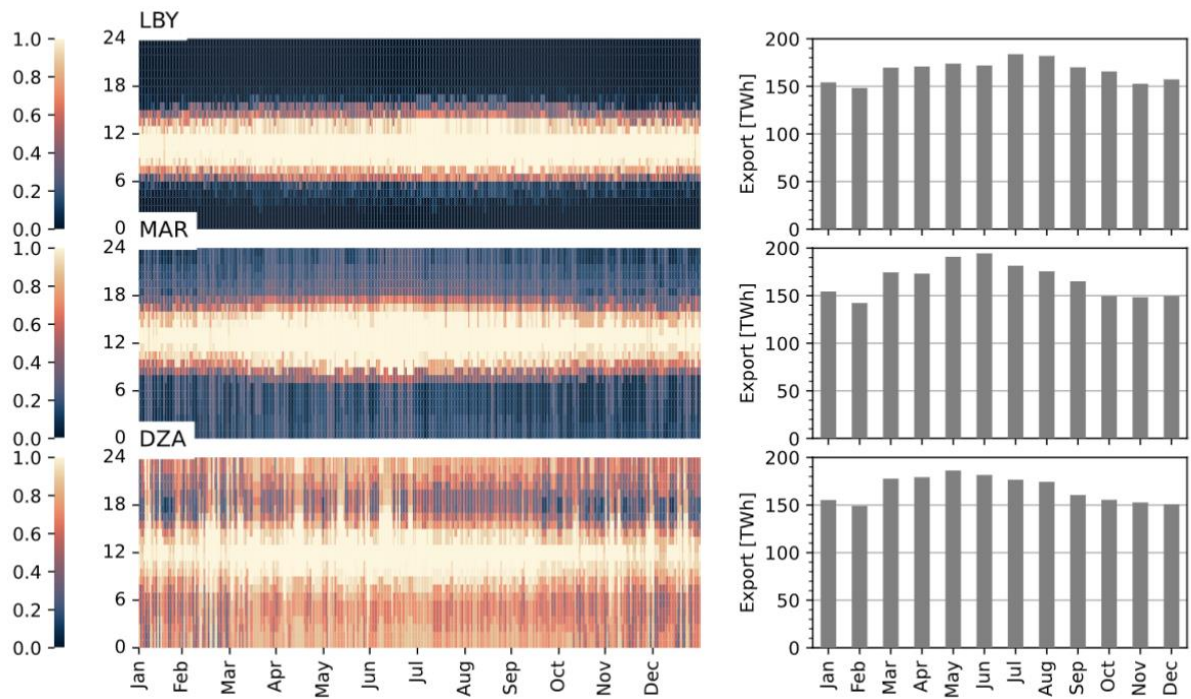

**Figure 9.** Seasonal hydrogen production profile for export with pipeline in Lybia, Morocco and Algeria for the weather year 2010 at export volume of 2000 TWh/a for the target year 2050.

For pipeline imports in the model, seasonal export profiles from the exporting countries are assumed, as it is assumed that hydrogen produced is exported directly to Europe without significant storage. Depending on the exporting country and the weather year, different seasonal profiles result, which are shown in Figure 9 as examples for the weather year 2010. The figure shows the seasonal fluctuations in hydrogen production. Higher hydrogen production can be observed in the summer months, which is due to higher solar radiation. In Morocco and Algeria, which use both PV and wind for hydrogen production, the months with the highest production are in early summer. In Libya, which primarily uses PV for hydrogen production, production is highest in midsummer. In addition, different profiles in daily production can be observed depending on the exporting country and the ratio of PV and wind in the system. In Libya, which primarily produces hydrogen from PV, hydrogen exports are limited to the daytime hours with high solar radiation. In Algeria, on the other hand, which uses both PV and wind for hydrogen production, hydrogen is also exported during the night. In summary, it can be said that the hydrogen export costs for liquid hydrogen in the countries and weather years considered are significantly higher than the export costs for pipeline export, with costs starting at 59€/MWh. The cost differences are primarily due to the additional costs of liquefying hydrogen. In countries with a high share of PV in hydrogen production, the range of export costs is lower due to the lower interannual fluctuations in PV generation. Furthermore, the analysis shows that North African countries have similar export costs and can therefore be substituted without a significant increase in costs.

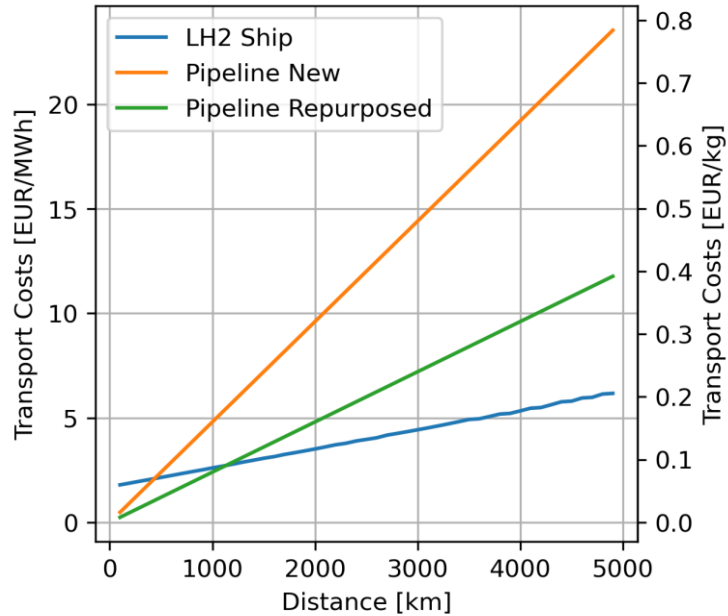

**Figure 10. Assumed cost of hydrogen transport per transport distance for liquid hydrogen transport and new and repurposed pipeline transport.**

Figure 10 shows the resulting transportation costs for hydrogen exports. For distances of less than 500 km for new pipelines and 1400 km for converted pipelines, the transportation of hydrogen by pipeline is cheaper than the transportation of liquid hydrogen by ship. For transportation distances of 1000km, transportation costs are estimated at 4.8 €/MWh for pipeline exports with new pipelines, 2.4 €/MWh for pipeline exports with repurposed pipelines and 6 €/MWh for ship exports. This means that the transportation costs for a distance of 1000 km account for around 4-8% of the total costs for pipeline transport and around 3% of the total costs for ship transport. Costs for regasification and the liquefaction of LH2 are not included in this graph. It should also be noted that the assumed transportation costs for liquid hydrogen ship transport are optimistic, as they do not include port charges, for example. The most favorable costs for hydrogen imports to Rotterdam result for liquid hydrogen imports from Western Sahara at 98 €/MWh. For pipeline imports to Spain, the most favorable costs are 63.5 €/MWh. Resulting hydrogen import costs for Spain are at 83 €/MWh and at 66.4 €/MWh for Italy for the weather year 2010. The costs found are similar to the costs found in the literature for hydrogen imports to Europe [7].

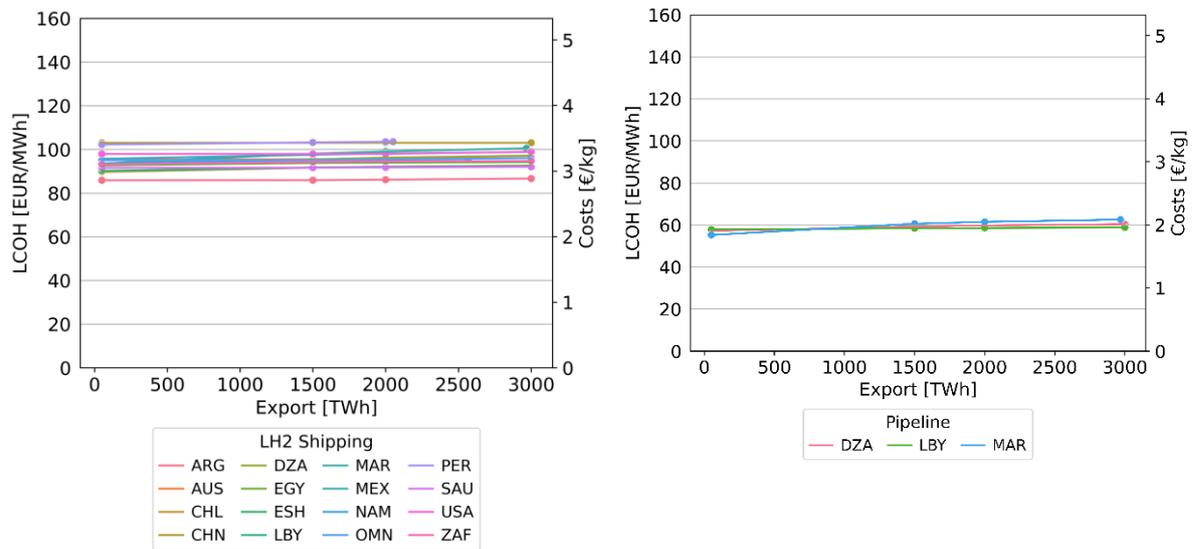

**Figure 11. Hydrogen export cost at different export volumes for ship and pipeline-based export in the year 2050 for the weather year 2010.**

## 6.2 Model Components

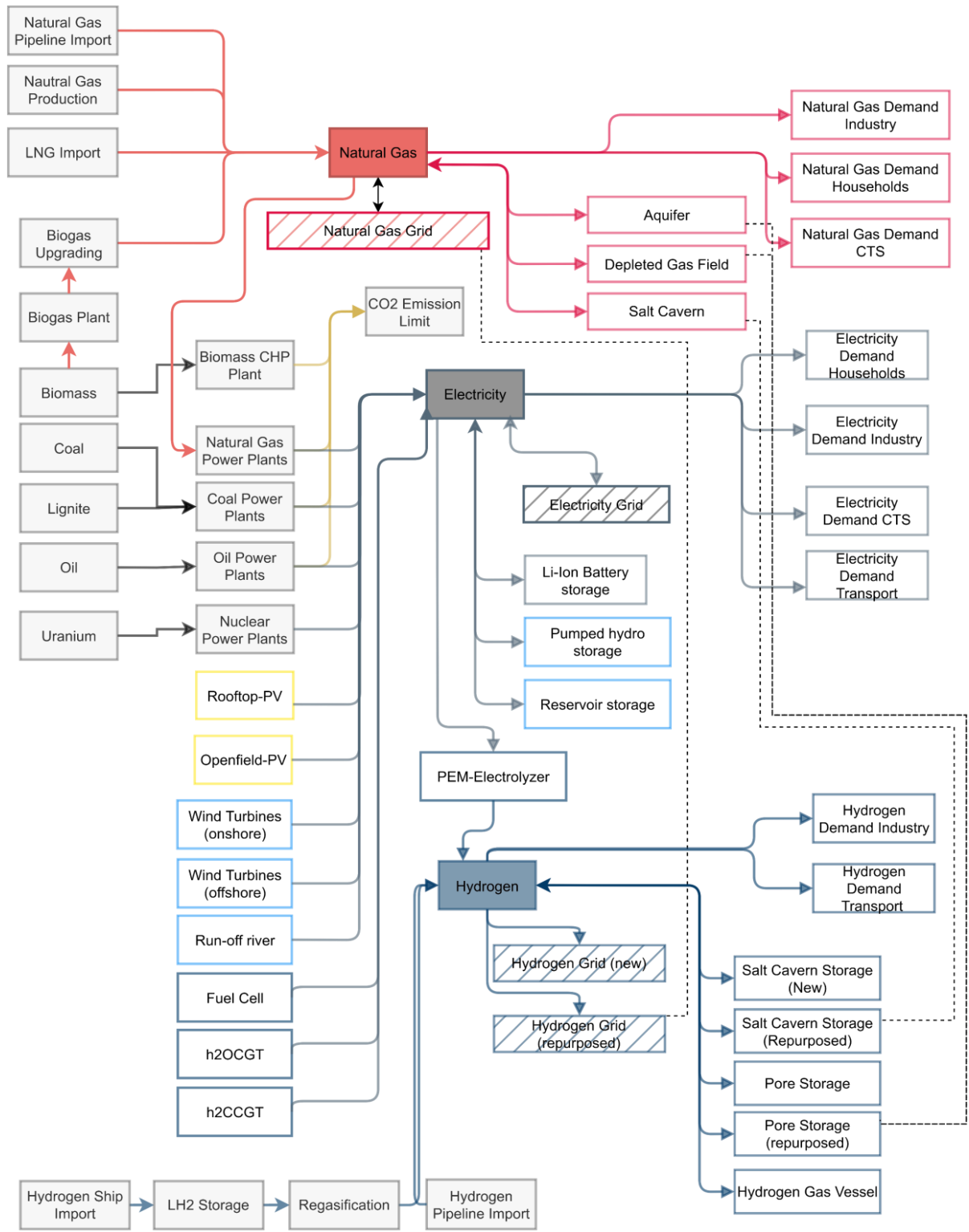

Figure 12. Block diagram of the resulting energy system model used for this investigation.

## 6.3 Commodity Demands

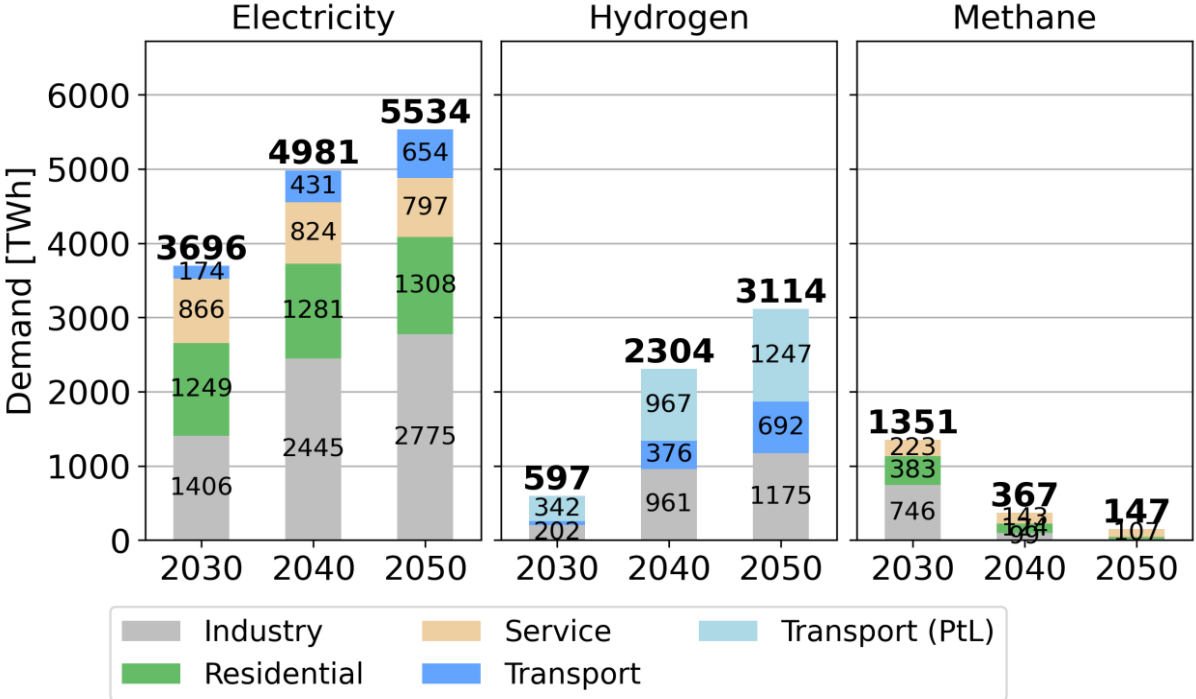

**Figure 13. Assumed electricity, hydrogen and methane demand in Europe in 2030, 2040 and 2050.**

Figure 13 shows the assumed electricity, hydrogen and methane demand in 2030, 2040 and 2050 in the model.

Electricity demand increases significantly due to electrification, reaching 2800 TWh in industry and 650 TWh in transport by 2050, driven by electric mobility. In households and services, electricity use remains stable, as growing heat pump deployment is offset by energy efficiency from building renovations. Total electricity demand rises from 3780 TWh in 2030 to 5000 TWh in 2040 and 5530 TWh in 2050.

Hydrogen demand is assumed only for industry and transport. By 2050, transport accounts for 1940 TWh—690 TWh for direct use and 1250 TWh for synthetic fuel production, including decarbonization of international aviation and shipping. Industrial hydrogen use reaches 1175 TWh, mainly for steel, chemicals, and high-temperature heat. Germany leads in hydrogen demand with 680 TWh in 2050, followed by France (398 TWh), the UK (395 TWh), Spain (256 TWh), and Italy (251 TWh). Total European demand rises from 600 TWh in 2030 to 2300 TWh in 2040 and 3110 TWh in 2050—slightly above literature averages due to inclusion of international transport.

Methane demand declines steadily, driven by electrification, hydrogen use, and energy efficiency, particularly in households and industry. In 2050, 150 TWh of residual methane demand remains due to the assumption that no conversion of natural gas use for cooking is taken into account.

## 6.4 ETHOS.FINE

The model developed for this study, ETHOS.fineEurope, is based on the open-source Python framework ETHOS.FINE (Framework for Integrated Energy System Analysis) [15]. ETHOS.FINE enables the modeling and optimization of multi-energy carrier systems with flexible spatial and temporal resolutions. The goal of the optimization is to minimize total annual system costs under technical and regulatory constraints. Core model components include:

- Sources: Represent energy generation or imports.
- Sinks: Represent demand or exports.
- Conversions: Capture energy conversion technologies (e.g., electrolysis, power plants).
- Storages: Model energy storage over time.
- Transmissions: Represent energy carrier transport between regions.

Time series aggregation can be applied using the tsam Python package to reduce computational complexity [16].

## 6.5 Energy Demand

Energy demand is calculated exogenously using stock models and tools that allow a scenario-based modeling of the development until 2050. No cost optimization was employed.

Due to the limited availability of datasets depicting future energy demand scenarios with high hydrogen shares across all sectors, and the absence of open-source tools for generating such scenarios, this study develops custom model-based demand projections. This section details the methodology for the residential and service sectors.

### 6.5.1 Residential and Service Sector Demand Model

Energy demand scenarios for the residential and service (covering commercial and public services) sectors are determined using a stock-flow model. It builds upon a comprehensive database of the building stock and its energy consumption, and considers various building types, ages, heating systems, end-uses (space heating, hot water, cooling, cooking, electrical appliances), renovation measures, and climatic conditions.

**Database Construction**

A custom database was constructed that characterizes the European building stock and energy use by fuel and end-use category (heating, hot water, cooling, appliances) at country level for EU-27 countries and UK.

**Primary data sources include:**

1. TABULA WebTool [69]: Provides archetypal building characteristics, including specific energy demand for heating and hot water, primary energy consumption, and GHG emissions for 17 European countries. It distinguishes buildings by construction year, type (e.g., single-family house (SFH), multi-family house (MFH), apartment block (AB)), installed heating and hot water systems, and offers renovation stages. Energy calculations use U-values (thermal transmittance coefficients of materials), country-specific Heating Degree Days (HDD), and system efficiencies.

2. JRC-IDEES [70]: Offers detailed Eurostat energy balance disaggregation (2000-2015) for EU27+UK, covering residential, service, transport, industry, and energy sectors. For buildings, it provides final energy consumption by end-use and energy carrier, new construction and renovation rates, and system distributions, alongside total floor area and electricity use for appliances.
3. Hotmaps database [71]: Developed for the Hotmaps project, it details final energy consumption for heating, cooling, and hot water by building type and age for EU-27+UK. It also provides floor area distributions by building category.

Building classes in the final database are defined by country (EU27+UK), building age (seven cohorts, e.g., <1945, 1945-1969, based on Hotmaps), building type (SFH, MFH, AB, from Hotmaps), heating system (e.g., gas, heat pump, hydrogen, from TABULA), hot water system (as heating, from TABULA), building renovation level (e.g., existing, light, nearly-zero energy buildings (NZEB), from TABULA), and heating system renovation level (e.g., existing, optimized, from TABULA).

**The merging methodology involves:**

- Specific energy consumption (per m²) for cooking, cooling, and electrical systems is derived from JRC-IDEES total floor area $A_{JRC}$, total energy consumption per process ($E_{process}$), and the relative market penetration of the system ($s_{process}$):
  - $SEC_{process\ per\ m^2} = \frac{E_{process}}{A_{JRC} * s_{process}}$
- For electrical appliances, $s_{process} = 1$ is assumed.
- Absolute floor areas for each building class $A_{class}$ are determined by combining JRC-IDEES total stock area and system distributions with Hotmaps' relative age/type distributions. A uniform distribution of heating systems across age classes is assumed due to data limitations.
- The number of buildings per class is $n_{building} = \frac{A_{class}}{A_{building,Tabula}}$ where $A_{building,Tabula}$ is the area of the TABULA reference building. The number of heated buildings or buildings with a specific process (e.g., cooling) is then calculated using respective shares from Hotmaps $s_{heated,Hotmaps}$ or JRC-IDEES $s_{heated,JRC}$.

For the service sector, lacking specific TABULA archetypes, MFH/AB building types serve as proxies. TABULA energy consumption values are normalized for the service sector using JRC-IDEES data for specific systems $E_{system,JRC}$ and age-class correction factors derived from TABULA archetypes.

The total energy consumption for each building class and process $E_{class,process}$ is then calculated as $E_{class,process} = n_{building,process} * A_{building} * SEC_{process\ per\ m^2}$.

For countries not in TABULA, characteristics from neighboring countries are used. TABULA calculations are performed for all building classes in each country using specific HDDs and characteristics.

Due to lack of certain heating systems in TABULA, some heating systems are substituted. Key assumptions include hydrogen boilers modeled as gas boilers, coal ovens as wood ovens, and geothermal systems as district heating systems. Further, manual mapping of TABULA to Hotmaps age classes is performed and proxy building types are used if SFH/MFH/AB data is missing in TABULA for a country.

Finally, calculated final energy consumption per m² for heating and hot water in the residential sector is calibrated to match total 2015 consumption figures from JRC-IDEES for each country:

$$SEC_{building,process,calibrated} = SEC_{building,process} * \frac{E_{JRC,process}}{E_{calc,process}}$$

Service sector demands are not recalibrated as they are directly derived from JRC-IDEES data without the use of TABULA energy consumption data.

**Model Description**

The building stock model follows a System Dynamics approach, adapting the methodology of [72], to simulate energy demand development from 2020 to 2050. It focuses on the impact of renovation and demolition measures on energy consumption for space heating, hot water, cooling, cooking, and electrical appliances, rather than techno-economic optimization or agent-based modeling. The database described in the previous section is updated at each annual time step.

The model comprises three main components: building stock, energy consumption, and weather conditions. Weather affects energy demand through country-level heating and Cooling Degree Days (CDD) for a specified historical year.

Energy consumption for cooling per m² is based on historical CDD, while cooking energy is assumed constant. Heating and hot water demands are building-type specific, calculated using the TABULA methodology with annual HDDs.

Building stock evolution is modeled annually:

1. Prioritization: Buildings are sorted by user-defined renovation priority, e.g., highest specific GHG emissions, highest specific energy consumption, or oldest age class.
2. Demolition: A target demolition area is met based on a user-specified annual rate, potentially excluding certain age classes or already renovated buildings.
3. Cooling System Adoption: The number of cooling systems increases based on a user-defined annual growth rate $d_{cool}$.
4. Renovations:
   Building Envelope: Deep renovations (to NZEB standard, TABULA level 3) and light renovations (improving TABULA level by one) occur based on user-specified annual rates and the renovation priority.
   Heating/Hot Water Systems: System replacement, e.g., gas boiler to heat pump, and system improvement, e.g., old gas boiler to modern equivalent, occur based on user rates. The distribution of new system types for replacements is user-defined. It is assumed that heating and hot water systems are replaced with the same technology type.
5. Population-Driven Stock Changes: Total required building area $A_{total}$ is calculated from projected population (UN data) assuming constant living area per capita. The difference between required and existing area drives new construction or additional demolition. New constructions meet NZEB standards (TABULA post-2010, renovation level 3) with a user-defined heating system distribution.

Model dynamics are controlled by user inputs such as the weather year for HDD and CDD determination, renovation/demolition rates, system technology choices, and prioritization strategies.

**Spatial and Temporal Disaggregation**

Annual country-level energy demands are disaggregated spatially to model-specific regions and temporally to an hourly resolution.

Spatial disaggregation for the residential sector uses population density data from WorldPop [73], assuming uniform distribution of building characteristics within a country. For the service sector, disaggregation is based on the number of employees per NUTS-2 region and sub-sector (Eurostat data [74]), and sub-sectoral floor area distributions (Hotmaps), largely following [75].

Temporal disaggregation to hourly resolution employs:

- when2heat tool [76]: (Reimplemented for performance) Generates dimensionless hourly load profiles for space heating, hot water, and cooling per region, building type, and process, based on weather data and building characteristics.
- Residential electricity: Load profiles from [77].
- Service sector electricity: Load profiles generated using the `demandlib` Python package [78].

### 6.5.2 Transport Sector Energy Demand

Energy demand in the transport sector was modeled using a previously developed methodology [79], distinguishing four modes: road, air, maritime, and rail, each subdivided into passenger and freight transport.

First, current (2019 baseline, unless otherwise specified) transport activity, in passenger-kilometers (pkm) and tonne-kilometers (tkm), was quantified at the NUTS-2 regional level. For road transport, passenger activity was derived by multiplying the NUTS-2 vehicle stock [80] with national average annual mileage per vehicle [70]. Road freight activity was aggregated from NUTS-3 tkm data [81]. Air transport activity was calculated from passenger or freight volumes between airports multiplied by the flight distance [82,83], with the resulting pkm/tkm distributed equally to the NUTS-2 regions of origin and destination airports. Maritime activity combined port-level passenger/freight throughput [82,84] with mode-specific distance coefficients (pkm/passenger or tkm/tonne) derived from [85] and [86]. National rail activity data [87,88] were disaggregated to NUTS-2 regions based on regional rail network density [89].

Second, transport activity was projected to 2050 using growth rates from the EU Reference Scenario 2020 [90] .

Third, this projected activity was allocated to various powertrain technologies based on future market share projections. These projections were sourced from literature where available [90–92] or developed by the author. The split of internal combustion engine vehicles between fossil and alternative fuels for road transport was an authorial assumption [79]. Market share development was assumed to be uniform across all European regions.

Fourth, energy consumption for each powertrain was calculated by combining the allocated transport activity with projected powertrain efficiencies, sourced from [33] and [86], assumed to be region-independent.

Finally, this energy consumption was converted into final energy carrier demands. Synthetic fuels (methanol, ammonia, kerosene, diesel, gasoline) were assumed to be produced from hydrogen and $CO_2$, and bio-based fuels were converted to biomass equivalents, with demands ultimately expressed as hydrogen equivalents where applicable. Electricity consumption (e.g., for electric vehicles) is used directly. Conversion factors for hydrogen to synthetic fuels were derived from Schemme et al. [93] and Bazanella et al. [94] and are given in [79]. For example, 1.306 kWh of hydrogen is required per kWh of Fischer-Tropsch fuels, and 1.24 kWh of hydrogen per kWh of gasoline via the methanol-to-gasoline route. Methanol and ammonia production require 1.139 kWh and 1.142 kWh of hydrogen per kWh of fuel, respectively.

Annual NUTS-2 regional energy demands were then temporally disaggregated to an hourly resolution. Following [33], hydrogen demand across all transport sectors was assumed to have a constant load profile. For electricity demand, specific load profiles from [33] were applied for road and rail transport, while a constant load profile was used for other modes.

### 6.5.3 Industry Sector Energy Demand

The evolution of energy carrier demand in the industry sector was modeled at the country level, drawing inspiration from system dynamics principles. While the EU Emissions Trading System (EU-ETS) [95] incentivizes decarbonization, this mechanism was not explicitly modeled. Instead, the adoption of alternative, low-emission production routes and the decarbonization of process heat were modeled using scenario-driven logistic functions. Energy efficiency improvements were not explicitly considered. The analysis covered industries and products primarily based on the IDEES database [70], with the JRC chemical industry database [96] used for greater detail in the chemical sector.

Model input preparation involved compiling five key datasets at the country level.

1. Current production volumes for relevant products were collated from IDEES, Eurostat/Prodcom [97], Rehfeldt et al. [98], Gumminski [99], and the JRC chemical industry database [96], using a defined prioritization to reconcile varying reference years and coverage. For the JRC chemical database, which provides capacities, national production volumes were disaggregated from EU28 totals using national capacity shares.
2. Alternative production routes (e.g., hydrogen direct reduction for steel) and process heat decarbonization options were defined for each IDEES process/product. For process heat decarbonization, a general assumption was made that process heat below 500°C could be supplied by electricity, while temperatures above 500°C would require hydrogen or biomass.
3. Process-specific heat demand temperature distributions were determined by mapping IDEES processes to data from Rehfeldt et al. [98] via EU-ETS activity IDs, yielding a percentage distribution of energy consumption across temperature levels.
4. Application-specific energy consumption (SEC) for feedstock, cooling, heating, and electricity, and associated energy carrier shares per tonne of product, were derived from the IDEES and JRC chemical databases. For the chemical industry, specific energy consumption data from the JRC database were used, while the energy carrier distribution for "basic chemicals" from IDEES was applied across all JRC chemical products.
5. Future production volumes were projected to 2050 using the PotenCIA dataset [100], which extends the IDEES database.

The evolution of energy consumption per IDEES process/product and country was simulated annually from a 2015 baseline (last available year in IDEES) to 2050. This simulation tracked scenario-based shifts in production routes using logistic functions and changes in energy carriers for process heat based on user-defined probabilities and target distributions for different temperature levels. If a production route changed, process heat sources were also assumed to change accordingly. For processes retaining conventional routes, heat sources could still transition independently. SEC for cooling and electricity applications, along with their energy carrier mix, were assumed constant. The simulation calculated the evolving SEC per tonne of product, which, combined with projected production volumes, yielded total annual energy carrier demands for each IDEES process/product and country.

Subsequently, country-level annual energy demands were spatially disaggregated to individual industrial plants. Plant-specific data, including location and production capacity/output (derived from allocated emissions and benchmarks using EU-ETS activity IDs), were compiled from the European Union Transaction Log (EUTL) [101] and the Swiss Pollutant Release and Transfer Register [102]. For steel and chemical sectors, where EUTL data lacked sufficient process detail, specific plant databases from Global Energy Monitor and Eurofer [103,104] (steel) and the JRC chemical industry database [96] were used. Energy consumption was allocated to plants based on their share of national production capacity for the respective IDEES process/product.

Finally, annual plant-level energy consumption was temporally disaggregated to an hourly resolution assuming a constant load profile for all industrial sectors, applications, and energy carriers, following [33].